\documentclass[conference]{IEEEtran}

\usepackage{tikz}
\usepackage{enumitem}
\usepackage{bbm}
\usepackage{amsthm}
\usepackage{amsmath}
\usepackage{url}
\usepackage[hidelinks]{hyperref}
\usepackage{versions}
\usepackage{dblfloatfix}
\usepackage{tablefootnote}
\usepackage{graphicx}
\usepackage{multirow}
\usepackage{subfigure}
\usepackage{colortbl}
\usepackage{xspace}
\usepackage{xcolor}
\usepackage{wasysym}

\newcommand{\method}{\textsc{SHAPr}\xspace}
\newcommand{\loo}{LOO\xspace}
\newcommand{\ml}{ML\xspace}
\newcommand{\iment}{$I_{ment}$\xspace}

\newcommand{\ilira}{$I_{lira}$\xspace}

\newcommand{\ioptimal}{$I_{optimal}$\xspace}
\newcommand{\iideal}{$I_{ideal}$\xspace}
\newcommand{\song}{\textsc{SPRS}\xspace}
\newcommand{\dtrain}{\protect{D_{tr}}\xspace}
\newcommand{\daux}{\protect{D_{aux}}\xspace}
\newcommand{\dtest}{\protect{D_{te}}\xspace}
\newcommand{\knn}{$K$-NN\xspace}
\newcommand{\mb}{\protect{$\mathcal{M}$}\xspace}
\newcommand{\adv}{\protect{$\mathcal{A}$}\xspace}
\newcommand{\mia}{\protect{MIA}\xspace}
\newcommand{\koh}{\protect{KIFS}\xspace}

\includeversion{full}
\excludeversion{short}  

\begin{document}

\title{\method: An Efficient and Versatile Membership Privacy Risk Metric for Machine Learning}

\author{
    \IEEEauthorblockN{Vasisht Duddu\IEEEauthorrefmark{1}, Sebastian Szyller\IEEEauthorrefmark{2}, N. Asokan\IEEEauthorrefmark{1}\IEEEauthorrefmark{2}}
    
    \IEEEauthorblockA{\IEEEauthorrefmark{1}University of Waterloo}
    \IEEEauthorblockA{\IEEEauthorrefmark{2}Aalto University}
    \IEEEauthorblockA{vasisht.duddu@uwaterloo.ca, contact@sebszyller.com, asokan@acm.org}
}

\maketitle

\begin{abstract}
Data used to train machine learning (ML) models can be sensitive.
Membership inference attacks (\mia{s}), attempting to determine whether a particular data record was used to train an ML model, risk violating \emph{membership privacy}.
ML model builders need a principled definition of a metric to quantify the membership privacy risk of (a) individual training data records, (b) computed independently of specific \mia{s}, (c) which assesses susceptibility to different \mia{s}, (d) can be used for different applications, and (e) efficiently. None of the prior membership privacy risk metrics simultaneously meet all these requirements.

We present \method, a membership privacy metric based on Shapley values which is a leave-one-out (\loo) technique, originally intended to measure the contribution of a training data record on model utility. We conjecture that contribution to model utility can act as a proxy for memorization, and hence represent membership privacy risk.

Using ten benchmark datasets, we show that \method is indeed effective in estimating susceptibility of training data records to \mia{s}.
We also show that, unlike prior work, \method is significantly better in estimating susceptibility to newer, and more effective \mia.
We apply \method to evaluate the efficacy of several defenses against \mia{s}: using regularization and removing high risk training data records.
Moreover, \method is versatile: it can be used for estimating vulnerability of different subgroups to \mia{s}, and inherits applications of Shapley values (e.g., data valuation).
We show that \method has an acceptable computational cost (compared to na\"{i}ve \loo), varying from a few minutes for the smallest dataset to $\approx$92 minutes for the largest dataset.

\end{abstract}

\section{Introduction}\label{sec:introduction}

As the use of sensitive data to build machine learning (\ml) models increases, assessing data privacy risks has become an important consideration. Several official reports from government institutions (NIST~\cite{nist}, the White House~\cite{whitehouse}, and the United Kingdom's Information Commissioner’s Office~\cite{ico}) have highlighted the importance of data privacy risk assessment.
Membership inference attacks (\mia{s}) are a potential threat to privacy of an individual's data used for training \ml models~\cite{mia,mlleaks,song2020systematic,yeom}.
These attacks infer whether a given data record was used to train that model.
For datasets containing an individual's sensitive data, \mia{s} constitute a privacy threat.
For instance, identifying that any individual's data was used to train a health-related \ml model may allow an adversary to infer the health status of that individual.
Hence, measuring the \emph{membership privacy risk} of training data records is essential for data privacy risk assessment.

Several existing tools, like MLPrivacyMeter~\cite{mlprivacymeter} and MLDoctor~\cite{liu2021mldoctor}, can quantify membership privacy risk.
They are based on measuring the success rate of known \mia{s}~\cite{mia}.
In addition, these attacks use \emph{aggregate metrics} such as accuracy, precision and recall over \emph{all} training data records, and are not designed for quantifying \emph{individual record-level} membership privacy risk. A record-level membership privacy risk metric allows the model builder to estimate the relative risk of different training data records. Additionally, it can help a user understand the privacy risk of contributing their data to the specific \ml task.

Song and Mittal~\cite{song2020systematic} proposed a record-level probabilistic privacy risk metric (hereafter referred as \song) defined as the likelihood of a data record being present in the target model's training dataset. \song is intended to be used by adversaries rather than model builders. Furthermore, it relies on a specific \mia's attack features to generate scores which can impact its effectiveness to assess susceptibility to newer and more effective \mia{s}.
Ideally, a membership privacy risk metric should capture the root cause of \mia{s}, namely, the memorization of training data records as suggested in prior work~\cite{mia}. Such a metric will be \textit{independent of any specific \mia} and thus be applicable to any future \mia{s} as well.  

A principled, attack-agnostic approach for estimating the memorization of training data records is to measure their influence on the model's utility. This can be done by using leave-one-out (\loo) training approach~\cite{feldman2020memorization,long2017measuring} where the influence of each data record is computed using the difference in model utility with and without that record in the training dataset. Long et al.~\cite{long2017measuring} proposed one such metric based on \loo computation which is independent of any specific attack.
However, directly using the na\"{i}ve \loo approach for each data record is computationally expensive~\cite{jia2019knn,jia19shapley,ghorbani19datashapley,jia2021scalability}. 

A good membership privacy metric must therefore be \emph{fine-grained} (measuring risk for individual records), \emph{attack-agnostic}, \emph{efficient}, and of course \emph{effective}. None of the existing metrics simultaneously satisfy all of these criteria.
We conjecture that Shapley values, a well-known notion in game theory used to quantify the contributions of individuals within groups~\cite{Shapley}, can fill this gap.
They approximate \loo computation to measure the influence of individual training data records on model utility~\cite{jia19shapley,ghorbani19datashapley}, thereby estimating the extent of their memorization.
Crucially, Shapley values can be \textit{efficiently} computed in one go for every training data record without having to \emph{train two models for each training data record} (with and without that data record in the training dataset) as typically done in na\"{i}ve \loo approach~\cite{jia2019knn,jia2021scalability}.

We make the following contributions.
\begin{enumerate}[leftmargin=*]
    \item We validate our conjecture by presenting \method, an \loo-based membership privacy risk metric using \emph{Shapley values}. \method is an \emph{attack-agnostic}, \emph{fine-grained} approach for estimating membership privacy risk for individual training data records.  (Section~\ref{sec:approach})
    \item We show that \method is \textit{effective} in assessing the susceptibility to  state-of-the-art \mia{s} across ten benchmark datasets. \method outperforms \song in assessing susceptibility to the most effective \mia. (Section~\ref{sec:miaeval}) 
    \item We demonstrate \method's applicability by showing that
    \begin{itemize}[leftmargin=*]
        \item it is effective at evaluating defences: \method can
        \begin{enumerate}[leftmargin=*]
            \item effectively capture the decrease in \mia accuracy on using defences like regularization (Section~\ref{sec:evaldef})
            \item show that removing high risk data records  does not necessarily reduce risk for the remaining data records (Section~\ref{sec:unlearning}); 
        \end{enumerate}        
        \item it is versatile: \method
        \begin{enumerate}[leftmargin=*]
            \item can be used to detect disparity in membership privacy risks across different sensitive subgroups (Section~\ref{sec:fairness}). 
            \item inherits other applications of Shapley values such as data valuation (Section~\ref{sec:dataval}).
        \end{enumerate}

    \end{itemize}
    \item We show that \method scores can be computed more \emph{efficiently} than the direct application of the \loo approach. (Section~\ref{sec:perfeval})
\end{enumerate}
\section{Background}\label{sec:background}

Consider a training dataset $\dtrain = \{x_i,y_i\}_{i=1}^n$ containing input features $x_i \in X$ and corresponding classification labels $y_i \in Y$ where $X$ and $Y$ are the space of all possible inputs and corresponding labels. An \ml classifier is a model $f_{\theta}$ which maps the inputs to the corresponding classification labels $f_{\theta}: X \rightarrow Y$. The function parameters $\theta$ are updated by minimizing the loss between the model's prediction $f_{\theta}(x)$ on input $x$ and the true labels $y$. The loss is minimized using training algorithms such as Stochastic Gradient Descent.


\subsection{Membership Inference Attacks}\label{sec:backmia}

\mia{s} exploit the difference in model behaviour on seen training data records and unseen test data records.
\mia{s} differentiate between members and non-members of the training dataset of a  model using the output predictions of that model, or some function of them. 

Shokri et al.~\cite{mia} proposed the first \mia that uses an \ml attack model to distinguish between a member and non-member based on the predictions of the target model. This was subsequently improved by several other papers which used different functions over model predictions to differentiate between membership and non-members: for instance, correction~\cite{yeom}, confidence~\cite{yeom,mlleaks}, entropy~\cite{song2020systematic,mia,mlleaks} and distance from decision boundary~\cite{label1,label2}.

In this work, we use modified entropy based attack~\cite{song2020systematic} and likelihood ratio based attack~\cite{carlini2021membership}; which are described in detail.

\noindent\textbf{Prediction Entropy}~\cite{song2020systematic,mia,mlleaks}. \adv may resort to a more sophisticated function defined over the set of confidence values in the prediction. The entropy in a model's prediction (i.e., information gain for \adv) is the uncertainty in predictions~\cite{mia,mlleaks}.
The entropy differs for training and testing data records which \adv can use as the basis for deciding whether an input data record was in the training set. For instance, the output for a training data record is likely to be close to a one-hot encoding, resulting in a prediction entropy close to zero. Testing data records are likely to have higher prediction entropy values. As with the previous method, \adv can choose a threshold for the prediction entropy to decide whether an input data record is a member or not.

A modification of prediction entropy attack was proposed by Song and Mittal~\cite{song2020systematic}.  
The prediction entropy is low for data records with both correct or incorrect classification predicted with high confidence by the model.
For a given data record $(x,y)$, the modified entropy function: $Mentr(f_{\theta}(x),y) = -(1 - f_{\theta}(x)_y) log(f_{\theta}(x)_y) - \sum_{i \neq y} (f_{\theta}(x)_i log(1 - f_{\theta}(x)_i))$,  accounts for this problem. Here, $f_{\theta}(x)_y$ indicates the prediction on record $x$ with correct label $y$. \adv thresholds the modified prediction entropy to determine the membership status: $I_{ment}(f_{\theta}(x),y)$ = $\mathbbm{1}\{Mentr(f_{\theta}(x),y) \leq \tau_y\}$. We refer to this \mia as \iment.

For \iment \mia, instead of using a fixed threshold of 0.5 over the prediction confidence as seen in original prediction entropy attack, the thresholds $\tau_y$ are adapted for each class to improve the \mia accuracy. This adaptive threshold gives the best \mia accuracy~\cite{song2020systematic}.


\noindent\textbf{Likelihood Ratio Test}~\cite{carlini2021membership}. Most prior \mia{s} focus on balanced accuracy, often resorting to reporting (high) true-positive rates at moderate false-positive rates. This is not meaningful for practical applications because at suitably low false-positive rates, their true-positive rates tend to be poor, thereby lowering the confidence in these \mia{s}. To address this, Carlini et al.~\cite{carlini2021membership} proposed an \mia based on the likelihood ratio test. The output predictions are scaled as $\rho(f(x)_y) = log(\frac{f(x)_y}{1-f(x)_y})$ followed by training multiple ``shadow models'' to estimate $\mathcal{N}(\mu_{in},\sigma_{in}^2)$ for members and $\mathcal{N}(\mu_{out},\sigma_{out}^2)$ for non-members. Here, $\mu_{in} and \sigma_{in}$ correspond to $\rho()$ of members while $\mu_{out}$ and $\sigma_{out}$ correspond to $\rho()$ of non-members. The membership of an arbitrary data record $x$ is predicted by measuring the likelihood of its loss under each of the distributions and return the membership corresponding to most likely distribution: $\dfrac{p(\rho(f(x)_y | \mathcal{N}(\mu_{in},\sigma_{in}^2))}{p(\rho(f(x)_y | \mathcal{N}(\mu_{out},\sigma_{out}^2))}$.
We refer to this \mia as \ilira.

\subsection{Song and Mittal's Privacy Risk Scores}\label{sec:song}

Song and Mittal~\cite{song2020systematic} describe a membership privacy risk metric (which we refer to as \song) that defines the membership privacy risk score of $z_i= (x_i,y_i)$ as the posterior probability that $z_i \in \dtrain$ given the output predictions from the model $f_{\theta}(x_i)$. 
They compute the score as $r(z_i) =P(z_i \in \dtrain | f_{\theta}(x_i))$. This probability is computed using Bayes' theorem as $\frac{P(z_i \in \dtrain) P(f_{\theta}(x_i) | z_i \in \dtrain)}{  P(z_i \in \dtrain) P(f_{\theta}(x_i) | z_i \in \dtrain) +  P(z_i \in \dtest) P(f_{\theta}(x_i) | z_i \in \dtest)}$.
They assume that the probability of the data record belonging to the training/testing dataset is equally likely, $P(z_i\in \dtrain )$ = $P(z_i \in \dtest)$ = 0.5. 
The membership privacy risk scores rely on training shadow models on $\daux$ to mimic the functionality of the target model. The conditional probabilities $P(f_{\theta}(x_i) | z_i \in \dtrain)$ and $P(f_{\theta}(x_i) | z_i \in \dtest)$ are then computed using the shadow model's output predictions on $\daux$'s training and testing dataset. Further, instead of using fixed threshold based prediction entropy \mia, each class has a threshold for deciding the data record's membership which are computed using $\daux$. The conditional probabilities are estimated per class $P(f_{\theta}(x_i) | z_i \in \dtrain)$ = \{$P(f_{\theta}(x_i) | z_i \in \dtrain, y=y_i)$\} across all class labels $y=y_i$.

Traditional \mia{s} require \adv to sample arbitrary data records to infer their membership status.
\song is designed as a tool for \adv to identify data samples which are more likely to be members instead of sampling a large number of data records.

\subsection{Memorization of Training Data in \ml}\label{sec:memorization}

Membership privacy risk (susceptibility to \mia{s}) occurs due the fact that \ml models, with their inherent large capacity, tend to ``memorize'' training data records~\cite{whitebox1,duddu2021gecko}. This results in distinguishable \ml model predictions on seen training data records and unseen testing data records~\cite{mia,mlleaks}.

A formal definition for ``memorization'' of a data record by an \ml model was proposed by Feldman~\cite{feldman2020memorization}. Memorization of $z_i$ can be estimated as the difference in the prediction of a model on input features $x_i$ when the model was trained with and without $z_i$ in its training set~\cite{feldman2020memorization}. Formally, for a specific model $f_{\theta}$ drawn from the set of models for a training algorithm $\mathcal{A}$, Feldman~\cite{feldman2020memorization} formulates memorization as follows: $mem(z_i, \dtrain,\mathcal{A}) = | Pr_{f_{\theta} \sim \mathcal{A}(\dtrain)}[f_{\theta}(x_i)=y_i]\\ - Pr_{f_{\theta} \sim \mathcal{A}(\dtrain\backslash z_i)}[f_{\theta}(x_i)=y_i]|$.
If $mem(z_i$, $\dtrain$, $\mathcal{A}$) is high, the model is likely to have memorized $z_i$. The above formulation of memorization is an \loo based approach which captures the extent to which the presence of a record in the training dataset \textit{influences} the model's output predictions~\cite{feldman2020memorization}. Feldman and Zhang~\cite{feldmanZhang} showed that memorization can be empirically estimated by computing the influence of data records to model utility.

To better understand the connection between memorization and membership privacy, we can think of membership privacy risk as follows: consider an \ml model is trained on $\dtrain$. \adv samples $z_i$ from $\dtrain$'s underlying data distribution where $z_i$= $(x_i,y_i)$ is the $i^{th}$ data record with input features $x_i$ and label $y_i$. 
\adv can query the model and observe the model's predictions (blackbox API access)~\cite{mia, song2020systematic, mlleaks} and parameters to compute intermediate layer output (whitebox access)~\cite{whitebox1,whitebox2}. \adv's goal is to infer whether $z_i$ $\in$ $\dtrain$ or $z_i$ $\not\in$ $\dtrain$. In practice, \adv can do this by estimating the \textit{influence} of $z_i$ on model's observables (predictions or intermediate layer output) after interacting with the \ml model.  Hence, measuring this \textit{influence} on the model observables acts as a signal for membership privacy risk for an individual data record $z_i$.

\subsection{Shapley Values}\label{sec:shapley}

An alternative approach to capture the influence of a training data record is by estimating Shapley values~\cite{ghorbani19datashapley,jia19shapley,jia2019knn,jia2021scalability}. Shapley values ($\phi_i$) are of the form,
\begin{equation}\label{shaporig}
    \phi_i =  \displaystyle \frac{1}{|\dtrain|} \sum_{\substack{S \subseteq \dtrain \backslash\{z_i\}}} \frac{1}{\binom{|\dtrain -1|}{|S|}} [U(S \cup \{z_i\}) - U(S)]
\end{equation}
where $S$ is a randomly chosen subset of
$\dtrain \backslash \{z_i\}$ and $U(S)$ (accuracy of $f_{\theta}$ on a testing dataset $\dtest$ when trained on $S$) is a utility metric. $\binom{|\dtrain-1|}{|S|}$ denotes the binomial coefficient for choosing $|\dtrain -1|$ elements from a set of $|S|$ elements.
Here, the Shapley value of $z_i$ is defined as the average marginal contribution of $z_i$ to $U(S)$ over all training data subsets $S \subseteq \dtrain {\backslash \{z_i\}}$.
Evaluating the Shapley function na\"{i}vely for all possible subsets with and without $z_i$ is computationally expensive (complexity of $O(2^{|\dtrain|}$ for $|\dtrain|$ data records~\cite{jia2021scalability}) and not scalable (leading to the same problem as with na\"{i}ve \loo)~\cite{feldman2020memorization,long2017measuring}. Note that computing Shapley values cannot be done by training $|\dtrain|+1$ models. Shapley value, by definition, require sampling a subset $S$ for which we train two models: one with and without $z_i$. This adds to the computationally complexity of na\"{i}ve \loo.

However, several prior work have proposed efficient algorithms which approximate the computation of Shapley values~\cite{jia19shapley,ghorbani19datashapley,jia2019knn,jia2021scalability}. We consider the most efficient algorithm in literature where Shapley values can be efficiently computed using a $K$-Nearest Neighbours (\knn) classifier as a surrogate model~\cite{jia2021scalability}.
Unlike the na\"{i}ve approach to computing Shapley values which requires training two models for \emph{each training data record}, the \knn model, once trained, can be used to compute the Shapley values for \emph{all training data records}. This improves the computational complexity to O($|\dtrain| log(|\dtrain|.|\dtest|)$) compared to exponential complexity of the formulation in Equation~\ref{shaporig}. 
We now outline this approach~\cite{jia2021scalability}.

For a given $z_i$, we can first compute the \emph{partial contribution} $\phi^{test}_i$ of a single test data record $z_{test}$ to the Shapley value $\phi_i$ of $z_i$, and then add up these partial contributions across the entire $\dtest$.

\noindent\textbf{Step 1: ``Sorting Phase''.} This phase of \knn classifier consists of passing $\dtrain$ and a single testing data record $z_{test} = (x_{test},y_{test}) \in \dtest$, as an input to the target classifier $f^l_{\theta}$ which is the output of the $l^{th}$ layer in the network. $f_{\theta}$ denotes final layer probability scores across all classes.
Following prior work on Shapley values~\cite{jia2019knn,jia2021scalability}, the outputs $f^1_{\theta}(\dtrain)$ and $f^1_{\theta}(x_{test})$ and their corresponding true labels are used for further computation.

\noindent\textbf{Step 2: ``Score Assignment''.} For $z_{test}$, the \knn classifier identifies the top $K$ closest training data records $(x_{\alpha_1},\cdots,x_{\alpha_K})$ with labels $(y_{\alpha_1},\cdots,y_{\alpha_K})$ using the distance between the predictions ($f^1_{\theta}(x_{\alpha_1}),\cdots,f^1_{\theta}(x_{\alpha_K})$) and $f^1_{\theta}(x_{test})$. We use $\alpha_j(S)$ to indicate the index of the training data record, among all data records in $S$, whose output prediction is the $j^{th}$ closest to $f^1_{\theta}(x_\text{test})$. For brevity, $\alpha_j(\dtrain)$ is written simply as $\alpha_j$. Following prior work on data valuation~\cite{jia2019knn,jia2021scalability}, we use $K=5$.

\noindent\textbf{Step 3.} The \knn classifier assigns majority label corresponding to the top $K$ training data records as
the label to $x_\text{test}$.
The probability of the classifier assigning the correct label is given as:
$P[f^1_{\theta}(x_\text{test}) = y_\text{test}]=\frac{1}{K}\sum_{i = 1}^{K} \mathbbm{1}[y_{\alpha_i} = y_\text{test}]$.
Hence, the utility of the classifier with respect to the subset $S$, and the single test data record  $z_{test}$, is computed as $U^{test}(S) =\frac{1}{K} \sum_{k=1}^{\min\{K,|S|\}} \mathbbm{1}[y_{\alpha_k(S)} = y_\text{test}]$.

\noindent\textbf{Step 4.}
Consider all the data records in $\dtrain$ after sorting as described above
$\{ \cdots, z_{\alpha_{i-1}}, z_{\alpha_i}, z_{\alpha_{i+1}}, \cdots \}$.
From Equation~\ref{shaporig}, the difference between the partial contributions for two \emph{adjacent} data records $z_{\alpha_i}, z_{\alpha_{i+1}} \in \dtrain$ is given by $\phi_{\alpha_i}^{test} - \phi_{\alpha_{i+1}}^{test} = \displaystyle \frac{1}{|\dtrain| -1} \sum_{\substack{S \subseteq \dtrain \backslash\{z_{\alpha_i},z_{\alpha_{i+1}}\}}} \frac{[U^{test}(S \cup \{z_{\alpha_i}\}) - U^{test}(S \cup z_{\alpha_{i+1}})]}{\binom{|\dtrain -2|}{|S|}}$

Using the \knn utility function: $U^{test}(S \cup \{z_{\alpha_i}\}) - U^{test}(S \cup z_{\alpha_{i+1}}) = \frac{\mathbbm{1}[y_{\alpha_i} = y_\text{test}] - \mathbbm{1}[y_{\alpha_{i+1}} = y_\text{test}]}{K}$.
Once the label for $x_{test}$ is assigned, the partial contribution can be computed recursively starting from the farthest data record: 

\begin{equation}\label{eqn:KNN_unweighted_class_last}
\phi^{test}_{\alpha_{|\dtrain|}}=\frac{\mathbbm{1}[y_{\alpha_{|\dtrain|}} = y_\text{test}]}{|\dtrain|}
\end{equation}

\begin{equation}\label{eqn:KNN_unweighted_class_rec}
\phi^{test}_{\alpha_i} = \phi^{test}_{\alpha_{i+1}}\!\! +  \frac{\mathbbm{1}[y_{\alpha_i} = y_\text{test}] - \mathbbm{1}[y_{\alpha_{i+1}} = y_\text{test}]}{K}
    \frac{\min\{K, i\}}{i}
\end{equation}

The fraction $\frac{\min\{K, i\}}{i}$ is obtained by simplifying the binomial coefficient (the full derivation can be found in Theorem 1 of Jia et al.~\cite{jia2019knn}).
The intuition behind Equation~\ref{eqn:KNN_unweighted_class_rec} is that the contribution of $z_{\alpha_i}$ is $0$ if the nearest neighbor of $z_{\alpha_i}$ in $S$ is closer to $z_{test}$ than $z_{\alpha_i}$, and $1$ otherwise.
Using the above steps, we get $\phi^{test}$ for each $z_{test}$ of size $\dtrain \times 1$.
This recursive formulation in Equation~\ref{eqn:KNN_unweighted_class_rec} can be extended across all $\dtest$ to obtain a matrix $[\phi^{test}_{i}]$ of size $\dtrain \times \dtest$.
The final Shapley values can be obtained by aggregating the partial contributions $\phi^{test}_{i}$ across $\dtest$.

\section{Problem Statement}\label{sec:problem}


We conjecture that Shapley values, by virtue of measuring influence on model utility, and hence the extent of memorization, can serve a good membership privacy risk metric by indicating the susceptibility of training data records to \mia{s}. Our goal is to verify this conjecture.
To this end, we lay out the system and adversary models (Section~\ref{sec:model}), describe the desiderata for designing such a metric (Section~\ref{sec:requirements}), and outline the limitations of prior work (Section~\ref{sec:limitations}). Finally, we discuss the challenges in evaluating membership privacy risk metric and how we address it (Section~\ref{sec:idealmetric}).

\subsection{System and Adversary Model}\label{sec:model}

\noindent\textbf{System Model.} We consider the perspective of a \mb who trains a model using a dataset contributed to by multiple participants. 
\mb wants to estimate the susceptibility of individual data records to \mia{s}. \mb has full access to the training ($\dtrain$) and testing ($\dtest$) datasets and can use them to compute membership privacy risk scores for each training data record.

\noindent\textbf{Adversary Model.} We describe the adversary model for the \mia{s}. The ground truth for the membership privacy risk metric for a given training data record is the degree to which an actual state-of-the-art \mia~\cite{mia,song2020systematic,yeom} succeeds against that record. We adapt the standard adversary model for \mia{s}~\cite{song2020systematic,mia} to \mb's perspective.

The standard adversary model from prior work~\cite{mia,song2020systematic} considers adversary \adv has access to the prediction interface of a model $f_{\theta}$ built using a training dataset $\dtrain$. \adv submits data records via the prediction interface and receives model outputs (this is a widely adapted setting for cloud-based \ml models in the industry). Given an input data record $x$, \adv can only observe the final output prediction $f_{\theta}(x)$. The \mia{s} considered use the full confidence vector~\cite{song2020systematic,mia} instead of the labels~\cite{label1,label2}. \adv does not know the underlying target model architecture and has access to an auxiliary dataset $\daux$ sampled from the same distribution as $\dtrain$. 

Prior \mia{s} assume partial overlap between \adv's $\daux$ and $\dtrain$~\cite{mia,song2020systematic}. 
However, we adapt the above adversary model to \mb's perspective. We assume that \adv's $\daux$ completely overlaps with $\dtrain$ which gives an upper bound on the membership privacy risk.
This is reasonable from \mb's perspective who has complete access to $\dtrain$ which is used to train the \ml model. This implies that \mia{s} which rely on shadow models (to learn the characteristics to differentiate between members/non-members) are directly using the target model for as the shadow models. In other words, the underlying target model architecture is known and used as shadow models. This setting corresponds to \mb simulating the strongest possible adversary with complete knowledge of $\dtrain$ who evaluates how accurate are \mia{s} by matching the \mia predictions with the ground truth membership status which is already known to \adv.

\subsection{Membership Privacy Metric: Requirements}\label{sec:requirements}

We identify the following requirements which should be satisfied while designing a membership privacy risk metric:

\begin{enumerate}[label=\textbf{R\arabic*},leftmargin=*]
    \item \label{req0} \textbf{Fine-grained.} The metric generates scores for measuring the membership privacy risk of individual training data records. This allows for a fine-grained membership privacy risk analysis of the training data records of an \ml model (Section~\ref{sec:approach}). 
    \item \label{req1} \textbf{Attack-Agnostic.} Ideally, the metric should capture the root cause of all \mia{s}, i.e., memorization of training data records by \ml models~\cite{mia,whitebox1,feldman2020memorization}. Hence, membership privacy risk scores resulting from the ideal metric must be computed independently of specific \mia{s}. This allows the scores to assess the membership privacy risks with respect to different \mia{s} (Section~\ref{sec:approach}). 
    \item \label{req2} \textbf{Effectiveness.} The membership privacy risk scores of training data records must correlate with the likelihood of success of \mia prediction against those records. This is computed using metrics such as F1 score, precision and recall computed between the scores after applying a threshold and \mia prediction (Section~\ref{sec:miaeval}).  Evaluation of effectiveness presumes the availability of a reliable ground truth for computing \mia predictions. We return to this consideration in Section~\ref{sec:idealmetric}.
    \item \label{req3} \textbf{Applicability.} The membership privacy risk scores, once computed, should be applicable to different use cases (Section~\ref{sec:applicability}). The metric should be effective to evaluate different defences against \mia{s} (Section~\ref{sec:applications}). Moreover, the versatility\footnote{Versatility is a design choice instead of a metric's property. Given two comparable techniques, the one having additional benefits is likely to be deployed.} of the metric to evaluate the susceptibility of sensitive subgroups to \mia{s} and estimating economic value. 
    \item \label{req4} \textbf{Efficiency.} Membership privacy risk scores resulting from the metric must be computed within a reasonable time and low computation overhead (Section~\ref{sec:perfeval}). 
\end{enumerate}

\subsection{Limitations of Existing Metrics}\label{sec:limitations}

Privacy assessment libraries such as MLPrivacyMeter~\cite{mlprivacymeter} and MLDoctor~\cite{liu2021mldoctor} quantify the membership privacy risk using existing \mia{s}.
They use aggregate metrics such as accuracy, precision and recall for \mia{s} across all training data records, and are not optimized for estimating the privacy risks of individual data records~\cite{song2020systematic}. Hence, such metrics do not satisfy the fine-grained requirement~\ref{req0}.

Song and Mittal propose \song which is a probabilistic membership privacy risk metric for individual data records~\cite{song2020systematic}.
The more effective an \mia is against a particular data record, the higher the score.
\song computes membership privacy risk scores for different training data records using \mia features for a specific \mia. For instance, \song, as indicated in the original paper, uses modified entropy over the output predictions from training and testing data records to compute the scores. This does not satisfy the attack-agnostic requirement~\ref{req1}.
We later show that \song does not satisfy the applicability requirement~\ref{req3} and is not effective for some of the applications (c.f. Section~\ref{sec:applicability}).

Long et al.~\cite{long2017measuring} propose Differential Training Privacy as a membership privacy metric based on the na\"{i}ve \loo approach: computing the difference between model predictions with and without a given training record in the $\dtrain$ and hence, the influence of that record on the model utility.
However, as we saw in Section~\ref{sec:shapley}, direct application of the \loo approach cannot scale to large datasets and models since it requires retraining the model to estimate the score for \emph{each} data record.  Hence, such a na\"{i}ve \loo approach does not satisfy the efficiency requirement~\ref{req4} (c.f. Section~\ref{sec:perfeval}).

\setlength\tabcolsep{2pt} 
\begin{table}[!htb]
\footnotesize
\caption{None of the prior metrics satisfy all the requirements.}
\begin{center}
\resizebox{0.45\textwidth}{!}{
\begin{tabular}{|l||c|c|c|}
\hline
\textbf{Requirements} & \textbf{MLPrivacyMeter~\cite{mlprivacymeter}} & \textbf{\song~\cite{song2020systematic}} & \textbf{Long et al.~\cite{long2017measuring}} \\
& \textbf{MLDoctor~\cite{liu2021mldoctor}} & &  \\
& (Attack Based) & (Probabilistic Metric) & (\loo Metric) \\
\hline
\ref{req0} \textbf{Fine-Grained} & $\Circle$   & $\CIRCLE$ & $\CIRCLE$  \\
\hline
\ref{req1} \textbf{Attack-Agnostic} & $\Circle$   & $\Circle$ & $\CIRCLE$  \\
\hline
\ref{req2} \textbf{Effectiveness} &  $\CIRCLE$ & $\CIRCLE$   & $\CIRCLE$ \\
\hline
\ref{req3} \textbf{Applicability} &  $\Circle$   & $\Circle$   & $\CIRCLE$  \\
\hline
\ref{req4} \textbf{Efficiency} &  $\CIRCLE$  & $\CIRCLE$  & $\Circle$   \\
\hline
\end{tabular}
}
\end{center}
\label{tbl:comparison}
\end{table}

Table~\ref{tbl:comparison} summarizes the prior work with respect to the different requirements that they satisfy. None of the prior work satisfy all the requirements for an ideal membership privacy risk metric. Since \loo metric based metric satisfies all but efficiency requirement, it begs the question of \textit{whether \loo metric can be improved to be an efficient and effective metric for estimating susceptibility of individual training data records to \mia{s}}. We focus on Shapley values due to availability of efficient algorithms in literature~\cite{jia19shapley,jia2019knn,jia2021scalability}.

\subsection{Challenges in Evaluating Effectiveness of Membership Privacy Risk Metrics}\label{sec:idealmetric}

To evaluate the effectiveness of a membership privacy risk metric we need  reliable ground truth.
One possible source of such a ground truth is an ideal \mia (\iideal) which predicts all training data records as members and non-training data records as non-members.
However, we argue that \iideal constitutes poor ground truth from the perspective of evaluating susceptibility to \mia{s} because, generally, $\dtrain$'s distribution is long-tailed~\cite{feldman2020memorization}. An \ml model generalizes well for records that appear frequently or are similar to each other, thereby allowing the model to learn a characteristic pattern over them. The model may simply memorize outliers that constitute the long tail of $\dtrain$'s distribution~\cite{feldman2020memorization}). An effective \mia can correctly predict the membership status of such \textit{memorized} training data records compared to those that the model has successfully generalized.


An alternative ground truth is an optimal \mia (\ioptimal) which predicts only the highly memorized training data records (i.e., with a higher influence on model predictions) as members and all remaining data records as non-members.
While the ground truth for \iideal is trivial, we cannot derive the ground truth for \ioptimal in the absence of a demonstrably optimal \mia. 
Hence, we can only evaluate membership privacy metrics with respect to specific \mia{s} rather than being able to assess susceptibility to \emph{any} \mia. The best we can do, therefore, is to assess effectiveness with respect to the best available \mia.

\section{\method: An \loo membership privacy risk metric}\label{sec:approach}

Shapley values, originally designed as a game-theoretic notion to quantify the contributions of individuals within groups to the utility of a given task~\cite{Shapley}, was previously proposed for data valuation~\cite{ghorbani19datashapley,ghorbani20distributional,jia19shapley,jia2019knn} and explainability~\cite{lundberg2017unified}.
In order to validate our conjecture that Shapley values are effective in estimating the membership privacy risk, we present \method, a membership privacy risk metric using Shapley values based on the algorithm in Section~\ref{sec:shapley}.

\method scores inherit certain properties from Shapley values which satisfy requirements in Section~\ref{sec:requirements}.
In the context of membership privacy risk, these properties can be formulated as follows:
\begin{enumerate}[label=\textbf{P\arabic*},leftmargin=*]
    \item \label{prop1} \textbf{Interpretable.} \method score ($\phi_i$) (Equation~\ref{shaporig}) of a data record $z_i = (x_i, y_i)$ is measured by how $z_i$'s addition to a training dataset $S$ influences utility $U()$ of the resulting model (Equation~\ref{shaporig}). Consequently, no influence (i.e., $U(S) = U(S \cup {z_i}$)) leads to a zero score for $z_i$. Similarly if two data records $z_i$ and $z_j$ have the same influence (i.e., $U(S \cup {z_i}) = U(S \cup {z_j}$), then they are assigned the same score.
    We can identify three ranges of \method scores that have associated semantics:

        \begin{enumerate}[leftmargin=*]
            \item \textbf{Case 1: $U(S \cup \{z_i\})$ = $U(S) \rightarrow \phi = 0$:} There is no difference in the model's output regardless of the presence of $z_i$ in the training dataset: $z_i$ has no membership privacy risk.
            \item \textbf{Case 2: $U(S \cup \{z_i\})$ $>$ $U(S)\rightarrow \phi > 0$:} $z_i$ contributed to increasing the model utility. Higher scores indicate higher likelihood of memorization which increases the susceptibility to \mia{s}.
            \item \textbf{Case 3: $U(S \cup \{z_i\})$ $<$ $U(S)\rightarrow \phi < 0$:} $z_i$ was harmful to the model's utility  
            (not learnt well by the model or is an outlier). It has a higher loss and is indistinguishable from testing data records which makes it less susceptible to \mia{s}.
        \end{enumerate}

        This clear semantic association allows us to set meaningful thresholds for \method scores that can be used to decide whether a data record is susceptible to \mia{s}. The natural choice for a threshold is zero, i.e., records with higher score are indicated as members due to higher model's memorization of those records.

    \item \label{prop2} \textbf{Additive.} $\phi_i$ is computed using
    $\dtest$ (Equation~\ref{shaporig}). Specifically, $\phi_i(U_k)$ represents the influence of $z_i$ on utility $U()$ w.r.t to $k^{th}$ testing data record. For two testing data records $k$ and $l$, $U_i(\{k,l\}) = U_i(k) + U_i(l)$. Hence, $\phi_i$ is the sum of the membership privacy risk scores of $z_i$ with respect to each testing data record. This property further implies \textit{group rationality}~\cite{ghorbani19datashapley,jia2019knn} where $U()$ is fairly and completely distributed amongst all the training data records.

    \item \label{prop4} \textbf{Heterogeneous.} Different training data records influence the model's utility differently and hence, have varying susceptibility to \mia{s} (referred to as ``heterogeneity"). \method assigns scores to training data records based on their individual influence on the model's utility. This is referred to as \emph{equitable distribution} of utility among the training data records in prior work~\cite{jia19shapley}.

\end{enumerate}

We will refer back to these properties while interpreting the results of our experiments (Sections~\ref{sec:miaeval} and~\ref{sec:applicability}).
By definition, \method, by virtue of using Shapley values, is fine-grained as it assigns scores for individual training data records based on their influence to model utility satisfying requirement~\ref{req0}.
Furthermore, the generation of \method scores do not use any \mia features required for performing \mia{s}. Hence, this makes \method an attack-agnostic metric, satisfying requirement~\ref{req1}.

\section{Experimental Setup}\label{sec:setup}

We systematically evaluate the effectiveness of \method using several datasets which are described in Section~\ref{sec:datasets}. We then describe the model architecture details for training on the datasets (Section~\ref{sec:arch}), and the metrics to evaluate the effectiveness of \method with respect to these \mia predictions used as a ground truth (Section~\ref{sec:metrics}). We finally describe the model utility on $\dtest$ and performance of different \mia{s} (Section~\ref{sec:summary}).

\subsection{Datasets}\label{sec:datasets}

We used ten datasets for our experiments. Following prior work~\cite{mia,song2020systematic}, we used the same number of training and testing data records from all the datasets for computing balanced accuracy for \mia{s}.

\noindent\textbf{\song Datasets.} Three datasets: TEXAS, LOCATION and PURCHASE, were also used to evaluate \song~\cite{song2020systematic} -- we refer to them as \song datasets. To facilitate comparison with \song, we used the same dataset partitions for the three \song datasets as described in \cite{song2020systematic}.

\noindent\textbf{LOCATION} contains the location check-in records of individuals. We used the pre-processed dataset from~\cite{mia} which contains 5003 data samples with 446 binary features corresponding to whether an individual has visited a particular location. The data is divided into 30 classes representing different location types. The classification task is to predict the location type given the location check-in attributes of individuals. As in prior work~\cite{song2020systematic}, we used 1000 training data records and 1000 testing data records. 

\noindent\textbf{PURCHASE} consists of shopping records of different users. We used a pre-processed dataset from~\cite{mia} containing 197,324 data records with 600 binary features corresponding to a specific product. Each record represents whether an individual has purchased the product or not. The data has 100 classes each representing the purchase style for the individual record. The classification task is to predict the purchase style given the purchase history. We used 19,732 train and test records as in prior work~\cite{song2020systematic}.

\noindent\textbf{TEXAS} consists of Texas Department of State Health Services' information about patients discharged from public hospitals. Each data record contains information about the injury, diagnosis, the procedures the patient underwent and some demographic details.
We used the pre-processed version of the dataset from~\cite{mia} which contains 100 classes of patient's procedures consisting 67,330 data samples with 6,170 binary features. The classification task is to predict the procedure given patient's attributes. We used 10,000 train and test records as in prior work~\cite{song2020systematic}.

\noindent\textbf{Additional Datasets.} We used seven other datasets: MNIST, FMNIST, USPS, FLOWER, MEPS, CREDIT and CENSUS. We rounded down the number of training data records in dataset to the nearest 1000 and split it in half between training and testing datasets. An exception to this is MNIST and FMNIST where we used the entire training dataset (60,000 data records) and testing dataset (10,000 data records) of different sizes to ensure the utility of the resulting model is sufficiently high.

\noindent\textbf{MNIST} consists of a training dataset of 60,000 images and a test dataset of 10,000 images that represent handwritten digits (0-9). Each data record is a 28x28 grayscale image with a corresponding class label identifying the digit. The classification task is to identify the handwritten digits. We used the entire training and testing set.

\noindent\textbf{FMNIST} consists of a training dataset of 60,000 data records and a test dataset of 10,000 data records that represent pieces of clothing. Each data record is a 28x28 grayscale image with a corresponding class from one of ten labels. The classification task is to identify the piece of clothing.

\noindent\textbf{USPS} consists of 7291 16x16 grayscale images of handwritten digits. There area total of 10 classes. 
The classification task is to identify the handwritten digits.
We used 3000 training data records and 3000 testing data records. 

\noindent\textbf{FLOWER} consists of 3670 images of flowers categorized into five classes---chamomile, tulip, rose, sunflower, and dandelion---with each class having about 800 320x240 images. The dataset was collected from Flickr, Google Images and Yandex Images.
The classification task is to predict the flower category given an image.
We used 1500 train and 1500 testing data records. 

\noindent\textbf{CREDIT} is an anonymized dataset from the UCI Machine Learning dataset repository which contains 30000 records with 24 attributes for each record. It contains information about different credit card applicants, including a sensitive attribute: the gender of the applicant.
There are two classes indicating whether the application was approved or not.
The classification task is to predict whether the applicant will default.
We used 15000 training data records and 15000 testing data records.

\noindent\textbf{MEPS} contains 15830 records of different patients that used medical services, and captures the frequency of their visits.
Each data record includes the gender of the patient, which is considered a sensitive attribute.
The classification task is to predict the utilization of medical resources as ``High'' or ``Low'' based on whether the total number of patient visits is greater than 10. We use 7500 training data records and 7500 testing data records.

\noindent\textbf{CENSUS} consists of 48842 data records with 103 attributes about individuals from the 1994 US Census data obtained from UCI Machine Learning dataset repository. It includes sensitive attributes such as gender and race of the participant.
Other attributes include marital status, education, occupation, job hours per week among others.
The classification task is to estimate whether the individual's annual income is at least 50,000 USD. We used 24000 training data records and 24000 testing data records.

We summarize the dataset partitions in Table~\ref{tab:dataset_summary}.

\begin{table}[!htb] 
\caption{Summary of dataset partitions for our experiments.}
\begin{center}
\footnotesize
\begin{tabular}{ |c|c|c| } 
 \hline
 \textbf{Dataset} & \textbf{Training Set Size} & \textbf{Testing Set Size} \\ 
 \hline
 \multicolumn{3}{|c|}{\textbf{\song Datasets}}\\
 \hline
 \textbf{LOCATION} & 1000 & 1000 \\ 
 \textbf{PURCHASE} & 19732 & 19732\\
 \textbf{TEXAS} & 10000 & 10000\\
 \hline
 \multicolumn{3}{|c|}{\textbf{Additional Datasets}}\\
 \hline
 \textbf{MNIST} & 60000 & 10000 \\ 
 \textbf{FMNIST} & 60000 & 10000 \\
 \textbf{USPS} & 3000 & 3000\\
 \textbf{FLOWER} & 1500 & 1500\\
 \textbf{MEPS} & 7500 & 7500 \\
 \textbf{CREDIT} & 15000 & 15000 \\
 \textbf{CENSUS} & 24000 & 24000 \\
 \hline
\end{tabular}
\end{center}
\label{tab:dataset_summary}
\end{table}

\subsection{Model Architecture}\label{sec:arch}

While the proposed \method scores are compatible with all types of machine learning models, we focus on deep neural networks in our evaluation.
We used a fully connected model with the following architecture: [1024, 512, 256, 128, $n$] with tanh() activation functions where $n$ is the number of classes.
This model architecture has been used in prior work on \mia{s}~\cite{mia,song2020systematic}.
\method is scalable to larger models such as ResNet (previously shown for data valuation for Shapley values~\cite{jia2021scalability,jia2019knn}) but we focus on model architectures used previously in privacy literature.

\subsection{Evaluation Metrics}\label{sec:metrics}

For all the experiments, we used accuracy of \mia{s} as the primary metric along with the average membership privacy risk score.


\noindent\textbf{Balanced Attack Accuracy} is the number of training and testing data records, of equal dataset sizes, which are correctly distinguished as members and non-members (reported in Table~\ref{tab:performance}). We also refer to this as simply ``attack accuracy''.

\noindent\textbf{Average membership privacy risk score} is the average over the membership privacy risk scores assigned to training data records by a metric to evaluate the membership privacy risk across a group of data records. 

As in prior work~\cite{song2020systematic}, we used three additional metrics to measure the success of the \method scores with respect to \iment and \ilira: precision, recall and F1 score.

\noindent\textbf{Precision} is the ratio of true positives to the sum of true positive and false positives.
This indicates the fraction of data records inferred as members which are indeed members.

\noindent\textbf{Recall} is the ratio of true positives to the sum of true positives and false negatives. This indicates the fraction of the training dataset’s members which are correctly inferred as members.

\noindent\textbf{F1 score} is the harmonic mean of precision and recall computed as $2\times\frac{precision \times recall}{precision + recall}$. The highest values is one indicates perfect precision and recall while the minimum value of zero is when either precision or recall are zero. ``Member'' is considered as a positive class.

\subsection{Summary of Model Utility and Attack Accuracy}\label{sec:summary}

We report the results obtained on training the target model in Appendix~\ref{app:accuracy}: Table~\ref{tab:performance} which presents the baseline test accuracy of target models trained with each dataset. For \song datasets, the performance obtained are similar to the results reported in Song and Mittal~\cite{song2020systematic}. We use their code\footnote{\url{https://github.com/inspire-group/membership-inference-evaluation/blob/master/privacy_risk_score_utils.py}} to generate attack performance as well as \song scores for training data records.
\section{Assessing the Effectiveness of \method}\label{sec:miaeval}
We begin by evaluating the effectiveness of \method (Requirement~\ref{req2}) by assessing how well \method scores correlate with the success of \mia{s}. We also compare \method and \song in terms of effectiveness. To facilitate this comparison, we first focus on \iment as the \mia providing the ground truth since it was used in the evaluation of \song in their original paper~\cite{song2020systematic}.
%
We threshold \method scores at zero (Section~\ref{sec:approach}).
For \song, we use $0.5$ as the threshold since it gives the best F1 score out of all the threshold  values (in $[0.5,1.0]$) tested in the original work~\cite{song2020systematic}.

For each dataset, we repeated the experiment ten times.
For each metric, we report the mean and standard deviation for the alignment with ground truth. 
To compare \method and \song, we start with the null hypothesis that both sets of results (representing the alignment of either metric with the ground truth) came from the same distribution. 
For  $p<0.05$ there is enough evidence to say that effectiveness of the metrics are not the same (i.e., one significantly outperforms the other).
Otherwise ($p \geq 0.05$), we do not have enough evidence to say that metrics perform differently.
We colour code the results: 1) \colorbox{orange!25}{orange}, the hypothesis cannot be rejected - \song and \method are comparable (similar mean and small standard deviation); 2) \colorbox{red!25}{red}, the hypothesis is rejected, \song outperformed \method 3); and \colorbox{green!25}{green}, the hypothesis is rejected, \method outperformed \song.

\noindent\textbf{\underline{Evaluation using F1-Score.}}
Following the approach used for \song~\ref{sec:song}, we first evaluate the effectiveness using F1 scores (Table~\ref{tab:f1score_iment}).

F1 scores of \method is well above 0.8 for all datasets, indicating that it is effective as a membership privacy metric.
However, we observe that \song outperforms \method on most datasets.
We conjecture that the potential reason for this seeming advantage of \song could stem from the fact that we use \iment as the source of ground truth while \song also uses the attack features from \iment to generate privacy risk scores.
Ideally, a membership privacy risk metric should be computed independently of specific \mia{s} (requirement~\ref{req1}). Such a metric is likely to be effective in estimating  susceptibility to any \emph{future} state-of-the-art \mia.
One way to evaluate this ``future-proofness'' is to repeat the comparative evaluation using a newer and more effective \mia as the source of the ground truth.

\setlength\tabcolsep{2pt}
\begin{table}[h]
\caption{Comparison of the effectiveness of \method and \song with respect to \iment using the F1 score. \colorbox{orange!25}{orange} indicates comparable results, \colorbox{red!25}{red} indicates \song outperforms \method and \colorbox{green!25}{green} indicates \method outperforms \song.}
\footnotesize
\begin{center}
\begin{tabular}{ |c|c|c| } 
\hline
  \textbf{Dataset} & \multicolumn{2}{c|}{\textbf{\iment}} \\ 
  & \song & \method \\
 \hline
 \multicolumn{3}{|c|}{\textbf{\song Datasets}}\\
  \hline
\textbf{LOCATION} & \cellcolor{red!25}0.94 $\pm$ 0.02 & \cellcolor{red!25}0.90 $\pm$ 0.02 \\
\textbf{PURCHASE} & \cellcolor{orange!25}0.89 $\pm$ 0.01 & \cellcolor{orange!25}0.89 $\pm$ 0.01 \\
\textbf{TEXAS}  & \cellcolor{red!25}0.95 $\pm$ 0.02 & \cellcolor{red!25}0.83 $\pm$ 0.01 \\
 \hline
 \multicolumn{3}{|c|}{\textbf{Additional Datasets}}\\
 \hline
 \textbf{MNIST}  & \cellcolor{green!25}0.72 $\pm$ 0.00 & \cellcolor{green!25}0.96 $\pm$ 0.00 \\
\textbf{FMNIST}  & \cellcolor{red!25}0.98 $\pm$ 0.00 & \cellcolor{red!25}0.94 $\pm$ 0.00 \\
\textbf{USPS}  & \cellcolor{orange!25}0.77 $\pm$ 0.15 & \cellcolor{orange!25}0.86 $\pm$ 0.10 \\
\textbf{FLOWER}  & \cellcolor{green!25}0.89 $\pm$ 0.00 & \cellcolor{green!25}0.96 $\pm$ 0.00 \\
\textbf{MEPS} & \cellcolor{red!25}0.96 $\pm$ 0.01 & \cellcolor{red!25}0.90 $\pm$ 0.04\\
\textbf{CREDIT} & \cellcolor{red!25}0.93 $\pm$ 0.03 & \cellcolor{red!25}0.89 $\pm$ 0.02 \\
\textbf{CENSUS} & \cellcolor{red!25}0.97 $\pm$ 0.02 & \cellcolor{red!25}0.90 $\pm$ 0.01 \\
\hline
\end{tabular}
\end{center}
\label{tab:f1score_iment}
\end{table}

\setlength\tabcolsep{2pt}
\begin{table}[h]
\caption{Comparison of the effectiveness of \method and \song with respect to \ilira using the F1 score. \colorbox{orange!25}{orange} indicates comparable results, \colorbox{red!25}{red} indicates \song outperforms \method and \colorbox{green!25}{green} indicates \method outperforms \song.}
\footnotesize
\begin{center}
\begin{tabular}{ |c|c|c| } 
\hline
  \textbf{Dataset} &
  \multicolumn{2}{c|}{\textbf{\ilira}}\\ 
  & \song & \method  \\
 \hline
 \multicolumn{3}{|c|}{\textbf{\song Datasets}}\\
  \hline
\textbf{LOCATION} & \cellcolor{green!25}0.87 $\pm$ 0.04 & \cellcolor{green!25}0.95 $\pm$ 0.01\\
\textbf{PURCHASE} & \cellcolor{orange!25}0.64 $\pm$ 0.14 & \cellcolor{orange!25}0.77 $\pm$ 0.20\\
\textbf{TEXAS}  & \cellcolor{green!25}0.69 $\pm$ 0.02 & \cellcolor{green!25}0.87 $\pm$ 0.01\\
 \hline
 \multicolumn{3}{|c|}{\textbf{Additional Datasets}}\\
 \hline
 \textbf{MNIST} &\cellcolor{green!25}0.78 $\pm$ 0.05 & \cellcolor{green!25}0.99 $\pm$ 0.01\\
\textbf{FMNIST} & \cellcolor{green!25}0.73 $\pm$ 0.04 & \cellcolor{green!25}0.84$\pm$ 0.02\\
\textbf{USPS} & \cellcolor{green!25}0.70 $\pm$ 0.03 & \cellcolor{green!25}0.93 $\pm$ 0.01\\
\textbf{FLOWER} & \cellcolor{orange!25}0.82 $\pm$ 0.03 & \cellcolor{orange!25}0.86 $\pm$ 0.04\\
\textbf{MEPS} & \cellcolor{orange!25}0.65 $\pm$ 0.07 & \cellcolor{orange!25}0.67 $\pm$ 0.06\\
\textbf{CREDIT} & \cellcolor{orange!25}0.62 $\pm$ 0.02 & \cellcolor{orange!25}0.64 $\pm$ 0.01\\
\textbf{CENSUS} & \cellcolor{green!25}0.16 $\pm$ 0.08 & \cellcolor{green!25}0.76 $\pm$ 0.02\\
\hline
\end{tabular}
\end{center}
\label{tab:f1score_ilira}
\end{table}

\noindent\textbf{\underline{Future-Proofness.}}
Recently Carlini et al.~\cite{carlini2021membership} proposed a new \mia \ilira.
They argued that a \mia which indicates some data record as vulnerable confidently is more effective than a \mia that does well on average.
\ilira was shown to be more effective than \iment as it has a higher true positive rate at a sufficiently low false positive rate~\cite{carlini2021membership}.

Therefore, we use \ilira as the ground truth\footnote{We use the keras implementation of \ilira \url{https://github.com/stanleykywu/model-updates} by the authors of \ilira.} to fairly compare \method and \song with respect to their future proofness. We find that \method significantly outperforms \song on most datasets, and is comparable on the remaining ones (Table~\ref{tab:f1score_ilira}).
This confirms our conjecture that \song's apparent advantage in Table~\ref{tab:f1score_iment} was due to the use of \iment for ground truth. We can thus conclude that \method, by virtue of being independent of any specific \mia, is an effective membership privacy risk metric which generalizes well to a newer, more effective \mia.

\noindent\textbf{\underline{Evaluation using recall.}}
Having shown that \method's outperforms \song in terms of F1 scores, we argue that for a membership privacy risk metric, recall is more important than precision. Failing to correctly identify a training data record at risk (false negative) is undesirable from a privacy perspective, whereas incorrectly flagging a record as risky (false positive) constitutes erring on the safe side.

Table~\ref{tab:recall} compares \method and \song using both \iment and \ilira as ground truth in terms of recall.
The recall for \method is close to perfect for \ilira, and outperforms \song across all the datasets.

\setlength\tabcolsep{2pt}
\begin{table}[h]
\caption{Comparing recall of \method and \song with respect to \iment and \ilira. \colorbox{orange!25}{orange} indicates comparable results, \colorbox{red!25}{red} indicates \song outperforms \method and \colorbox{green!25}{green} indicates \method outperforms \song.}
\begin{center}
\footnotesize
\begin{tabular}{ |c|c|c|c|c| } 
\hline
  \textbf{Dataset} & \multicolumn{2}{c|}{\iment \textbf{Recall}} & \multicolumn{2}{c|}{\ilira \textbf{Recall}}  \\ 
  & \song & \method & \song & \method \\
 \hline
 \multicolumn{5}{|c|}{\textbf{\song Datasets}}\\
  \hline
\textbf{LOCATION} & \cellcolor{red!25}0.95 $\pm$ 0.02 & \cellcolor{red!25}0.87 $\pm$ 0.01 & \cellcolor{green!25}0.81 $\pm$ 0.06 & \cellcolor{green!25}0.97 $\pm$ 0.02\\
\textbf{PURCHASE} & \cellcolor{orange!25}0.82 $\pm$ 0.02 & \cellcolor{orange!25}0.81 $\pm$ 0.01 & \cellcolor{green!25}0.64 $\pm$ 0.09 & \cellcolor{green!25}0.98 $\pm$ 0.00 \\
\textbf{TEXAS} & \cellcolor{red!25}0.96 $\pm$ 0.01 & \cellcolor{red!25}0.73 $\pm$ 0.03 & \cellcolor{green!25}0.60 $\pm$ 0.03 & \cellcolor{green!25}0.89 $\pm$ 0.01 \\
 \hline
 \multicolumn{5}{|c|}{\textbf{Additional Datasets}}\\
 \hline
 \textbf{MNIST} & \cellcolor{green!25}0.57 $\pm$ 0.01 & \cellcolor{green!25}0.94 $\pm$ 0.00 & \cellcolor{green!25}0.64 $\pm$ 0.07 & \cellcolor{green!25}0.99 $\pm$ 0.00 \\
\textbf{FMNIST} & \cellcolor{red!25}0.98 $\pm$ 0.03 & \cellcolor{red!25}0.89 $\pm$ 0.03 & \cellcolor{green!25}0.71 $\pm$ 0.08 & \cellcolor{green!25}0.99 $\pm$ 0.00 \\
\textbf{USPS} & \cellcolor{green!25}0.76 $\pm$ 0.07 & \cellcolor{green!25}0.98 $\pm$ 0.01 & \cellcolor{green!25}0.58 $\pm$ 0.04 & \cellcolor{green!25}1.00 $\pm$ 0.00 \\
\textbf{FLOWER} & \cellcolor{green!25}0.81 $\pm$ 0.04 & \cellcolor{green!25}0.94 $\pm$ 0.01 & \cellcolor{green!25}0.86 $\pm$ 0.08 & \cellcolor{green!25}1.00 $\pm$ 0.00 \\
\textbf{MEPS}  & \cellcolor{red!25}0.96 $\pm$ 0.01 &  \cellcolor{red!25}0.91 $\pm$ 0.01 & \cellcolor{green!25}0.91 $\pm$ 0.05 & \cellcolor{green!25}0.98 $\pm$ 0.01\\
\textbf{CREDIT} & \cellcolor{red!25}0.98$\pm$ 0.05 & \cellcolor{red!25}0.92 $\pm$ 0.02 & \cellcolor{green!25}0.93 $\pm$ 0.06 & \cellcolor{green!25}0.99 $\pm$ 0.00\\
\textbf{CENSUS} & \cellcolor{red!25}0.99 $\pm$ 0.00 &  \cellcolor{red!25}0.87 $\pm$ 0.02 & \cellcolor{green!25}0.14 $\pm$ 0.08 & \cellcolor{green!25}0.97 $\pm$ 0.01 \\
\hline
\end{tabular}
\label{tab:recall}
\end{center}
\end{table}

No membership privacy risk metric can be equally effective against \emph{all} \mia{s}.
An optimal metric is likely to perform better with respect to more effective \mia{s}, than with less effective ones. Given that \method performs better on the more effective \mia (\ilira) than the less effective one (\iment), we suggest that \method is the better metric.
\section{Applicability of \method}\label{sec:applicability}

We evaluate \method in terms of its applicability (Requirement~\ref{req3}). 
First, we show how \method can be used to evaluate defences (Section~\ref{sec:applications}) followed by evaluating the versatility of \method (Section~\ref{sec:versatility}).
In this section, we revert to using \iment as the source for the ground truth because \method fares worse on \iment compared to \song. Hence, choosing \iment gives the least advantage to \method.
Recall that that \method is still effective when using \iment as ground truth (Tables ~\ref{tab:f1score_iment} and \ref{tab:recall}).

\subsection{Using \method to Evaluate Defences}\label{sec:applications}

Having shown that \method can effectively assess susceptibility to \mia{s}, we use \method to evaluate different potential defences that \mb can deploy: 1) using regularization-based defences (Section~\ref{sec:evaldef}) and, 2) retraining the model after removing vulnerable training data records (Section~\ref{sec:unlearning}).

\subsubsection{Evaluation of L2 Regularization}\label{sec:evaldef}

Prior work has shown that L2 regularization can be used as a defence against \mia{s}~\cite{l2def}.
Specifically, the average \method scores across all training data records should decrease when an effective defense is deployed. 
Following the experiment setup used for the evaluation of \song~\cite{song2020systematic}, we consider the \song datasets, namely, LOCATION, PURCHASE and TEXAS.

\begin{figure}[!htb]
    \centering
    \includegraphics[width=0.55\columnwidth]{{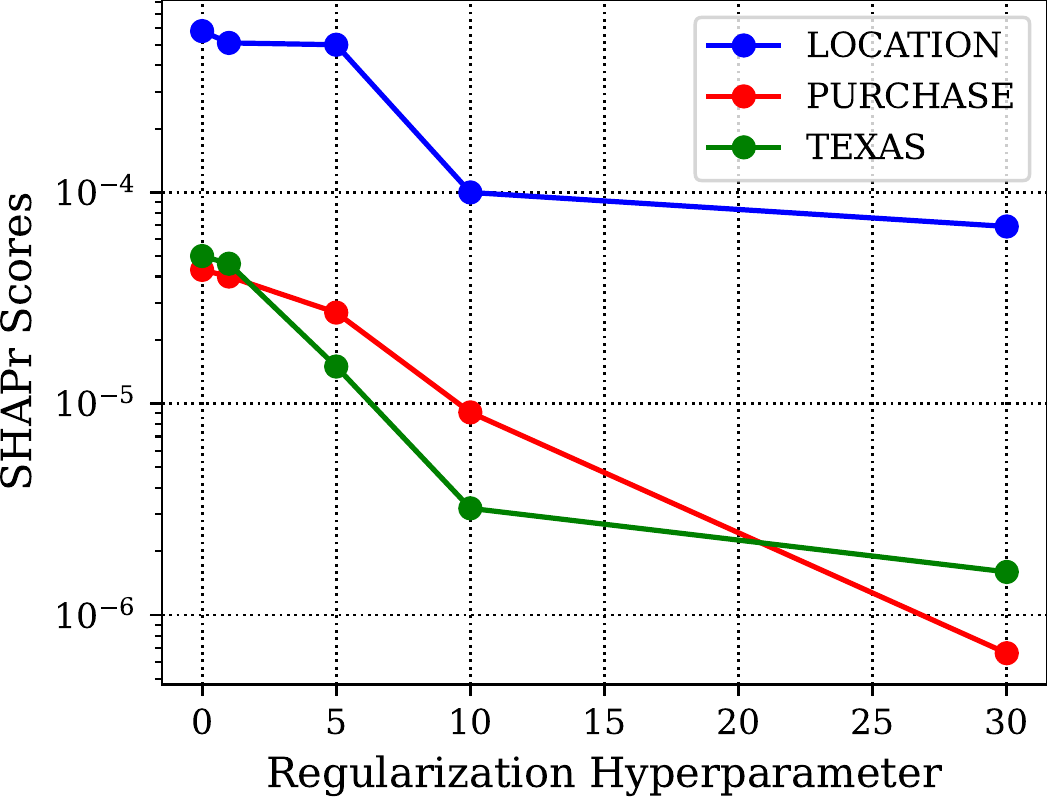}}
    \caption{Visual trend shows \method scores decrease on increasing the regularization hyperparameter.}
    \label{fig:l2reg}
\end{figure}

We compute the average \method scores for all training data records to see the trend of average privacy risk with increasing regularization hyperparameter.
In Figure~\ref{fig:l2reg}, we can see that \method scores decrease on increasing the regularization.

\subsubsection{Impact of Data Removal}\label{sec:unlearning}

In data valuation research, it is well-known that removing records with high Shapley values will harm the utility of the model, and removing records with low values will improve it~\cite{jia19shapley,jia2021scalability}.
Hence, it begs the question whether removal of records with high \method scores improves the membership privacy risk of a dataset, by reducing its overall susceptibility to \mia{s}. This has been explored as a possible defence in prior work as well~\cite{carlini2022privacy,long2017measuring}.
To verify whether \method can measure the effectiveness of the defence, we removed a fraction (up to $50$\%) of records with the highest \method scores.
Also, we randomly removed testing data records so as to keep the same number of member and non-member records as in previous experiments.
Following Section~\ref{sec:evaldef}, we consider the \song datasets: LOCATION, PURCHASE and TEXAS.

\begin{figure}[!htb]
\centering
\includegraphics[width=0.55\columnwidth]{{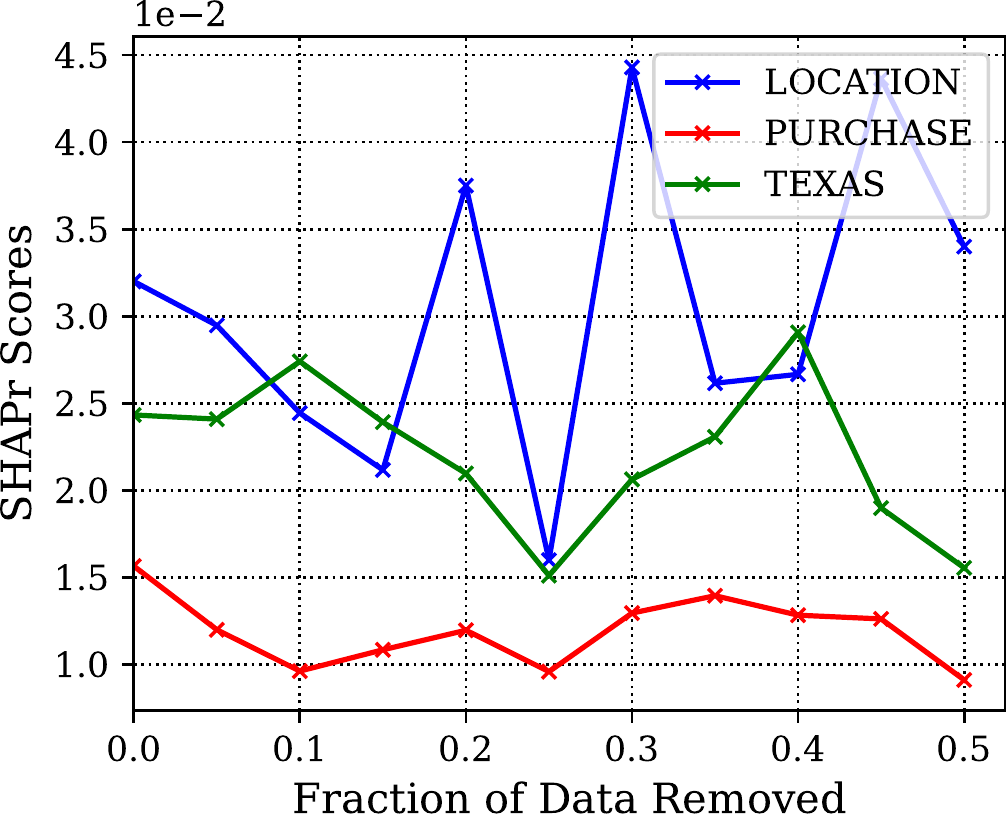}}
\caption{Removing a fraction of training data records with high \method scores does not reduce the risk for the remaining records.}
\label{fig:scores_interpret_high}
\end{figure}

Figure~\ref{fig:scores_interpret_high} summarizes the results.
Removing an increasing number of records with high \method scores does not necessarily reduce the membership privacy risk for the remaining records.
No consistent upward (or downward) trend was visible for the scores of the remaining records.
Interestingly, depending on the number of removed samples, the scores fluctuate.
A possible explanation is that once risky data records are removed, and a new model is trained using the remaining records. The influence of remaining records to the revised model and their memorization changes, thereby changing their \method scores. This matches with the observation by prior work~\cite{carlini2021membership,long2017measuring}.

A similar result was observed in prior work~\cite{long2017measuring}. However, Long et al.'s~\cite{long2017measuring} experiment was limited to minimal removal (only 20 records from 1.6 million records which is $<1$\%).
Furthermore, their analysis used $1.6$ million training data records with a Naive Bayes classifier rather than with a deep neural network. For a neural network, the computation of their scores on large datasets would be intractable (c.f. Section~\ref{sec:perfeval}).
With \method, we are able to confirm that this observation holds broadly across more complex deep neural networks and for a larger number of removed records (up to $50$\% vs. $<1\%$). 

\subsection{Versatility of \method}\label{sec:versatility}

To underscore the versatility of \method, we describe two further applications of \method, enabled thanks to the properties of Shapley values.
We show that \method can estimate the disparity of membership privacy risk across sensitive subgroups (Section~\ref{sec:fairness}), and we discuss \method{'s} applicability for data valuation (Section~\ref{sec:dataval}). 

\subsubsection{Privacy Risk of Sensitive Subgroups}\label{sec:fairness}
Prior work has shown that different subgroups with sensitive attribute (e.g., race or gender) have disparate vulnerability to membership inference attacks (\mia{s})~\cite{yaghini2019disparate}.
We evaluated whether \song and \method can correctly identify this disparity.

\begin{figure}[!htb]
    \centering
    \subfigure[CENSUS (race)]{\includegraphics[width=0.23\textwidth]{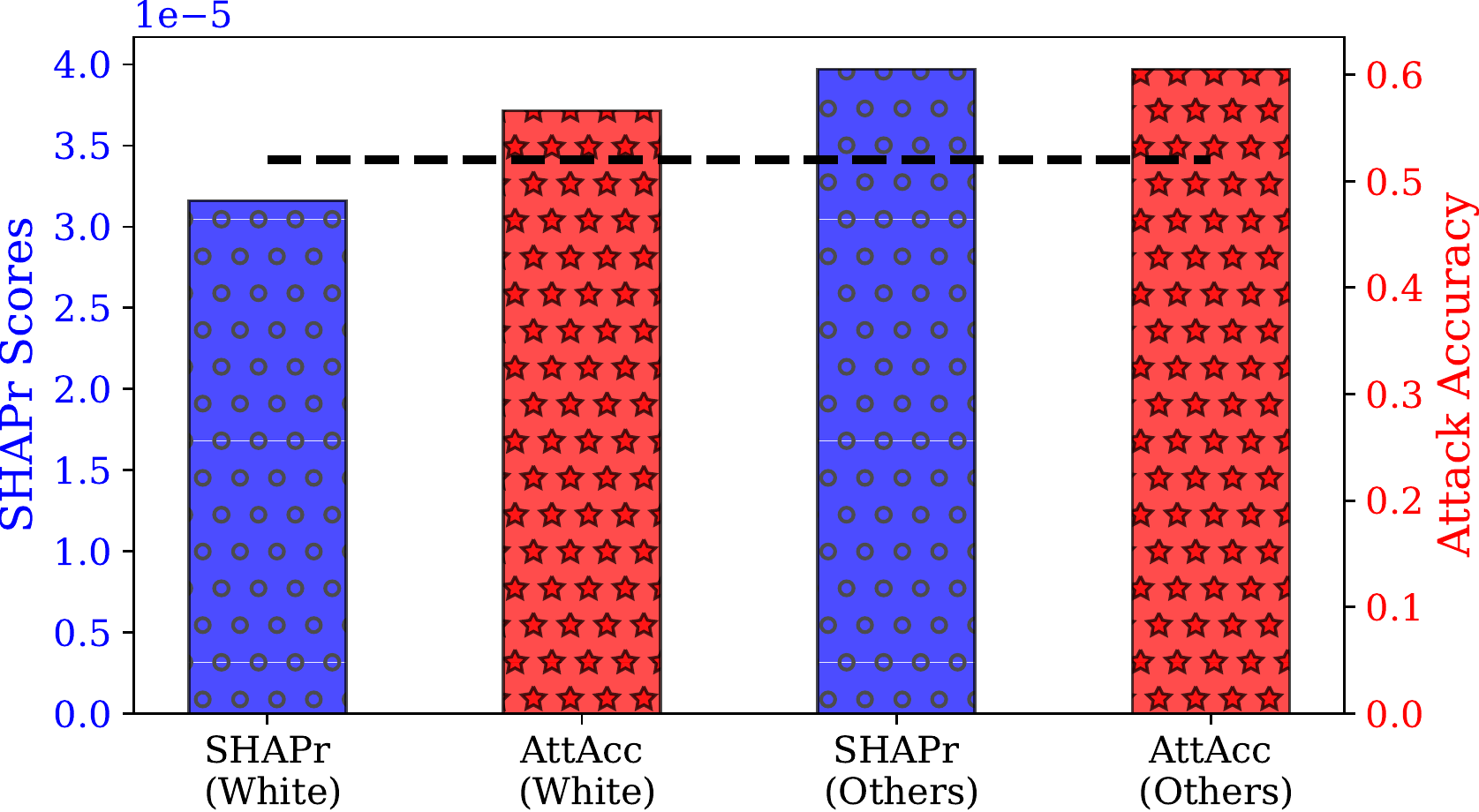}}
    \subfigure[CENSUS (gender)]{\includegraphics[width=0.23\textwidth]{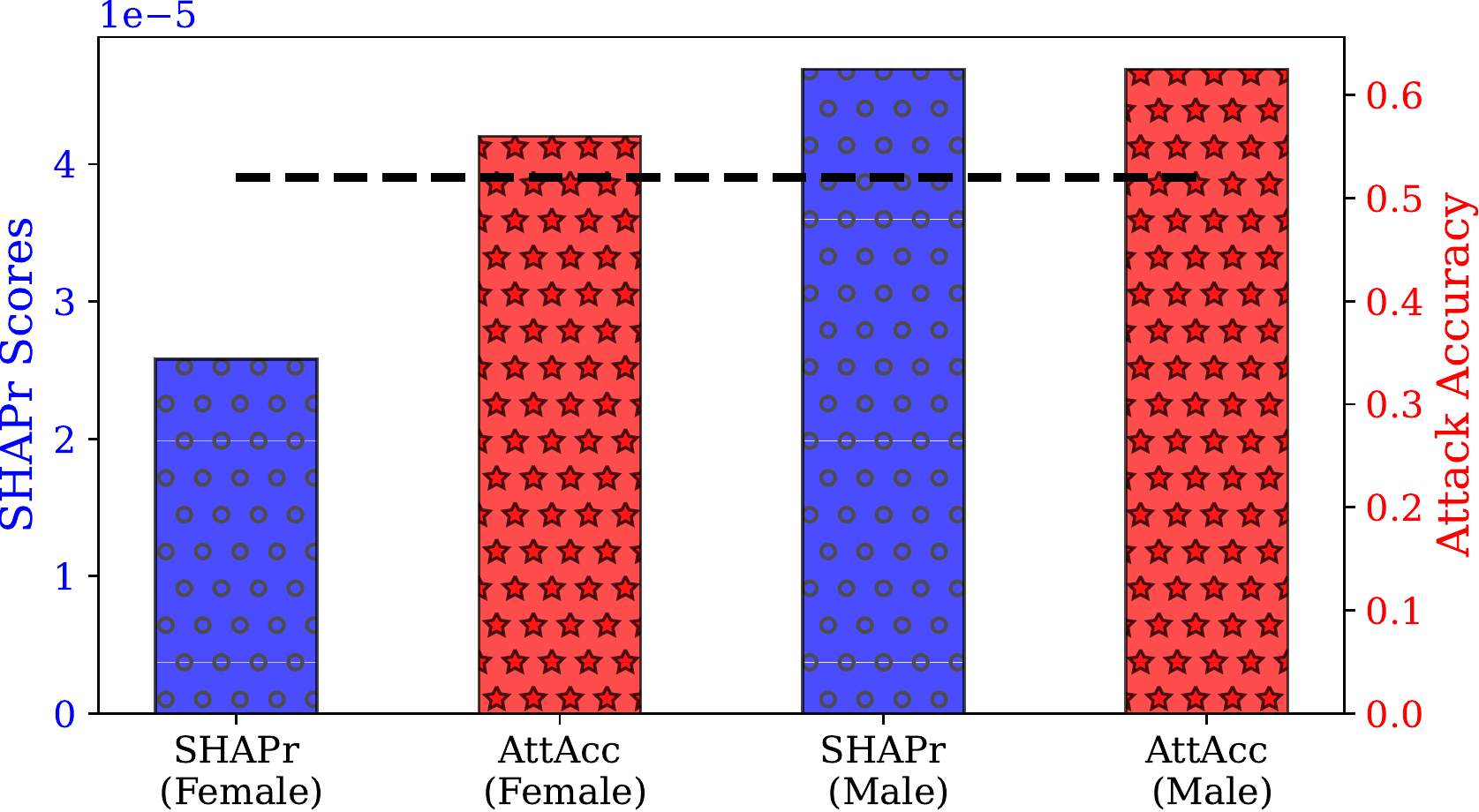}}\\
    \subfigure[CREDIT (gender)]{\includegraphics[width=0.23\textwidth]{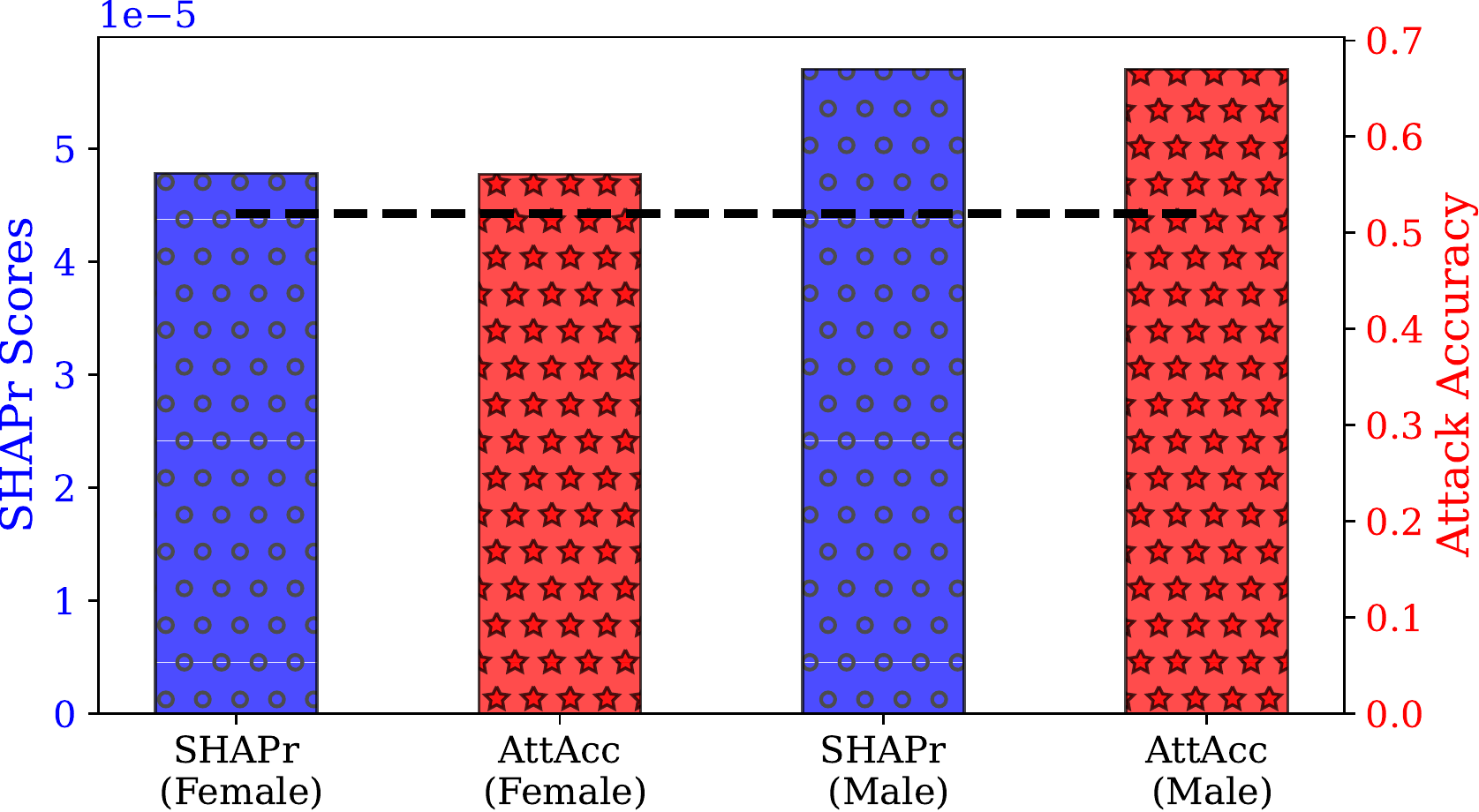}}
    \subfigure[MEPS (gender)]{\includegraphics[width=0.23\textwidth]{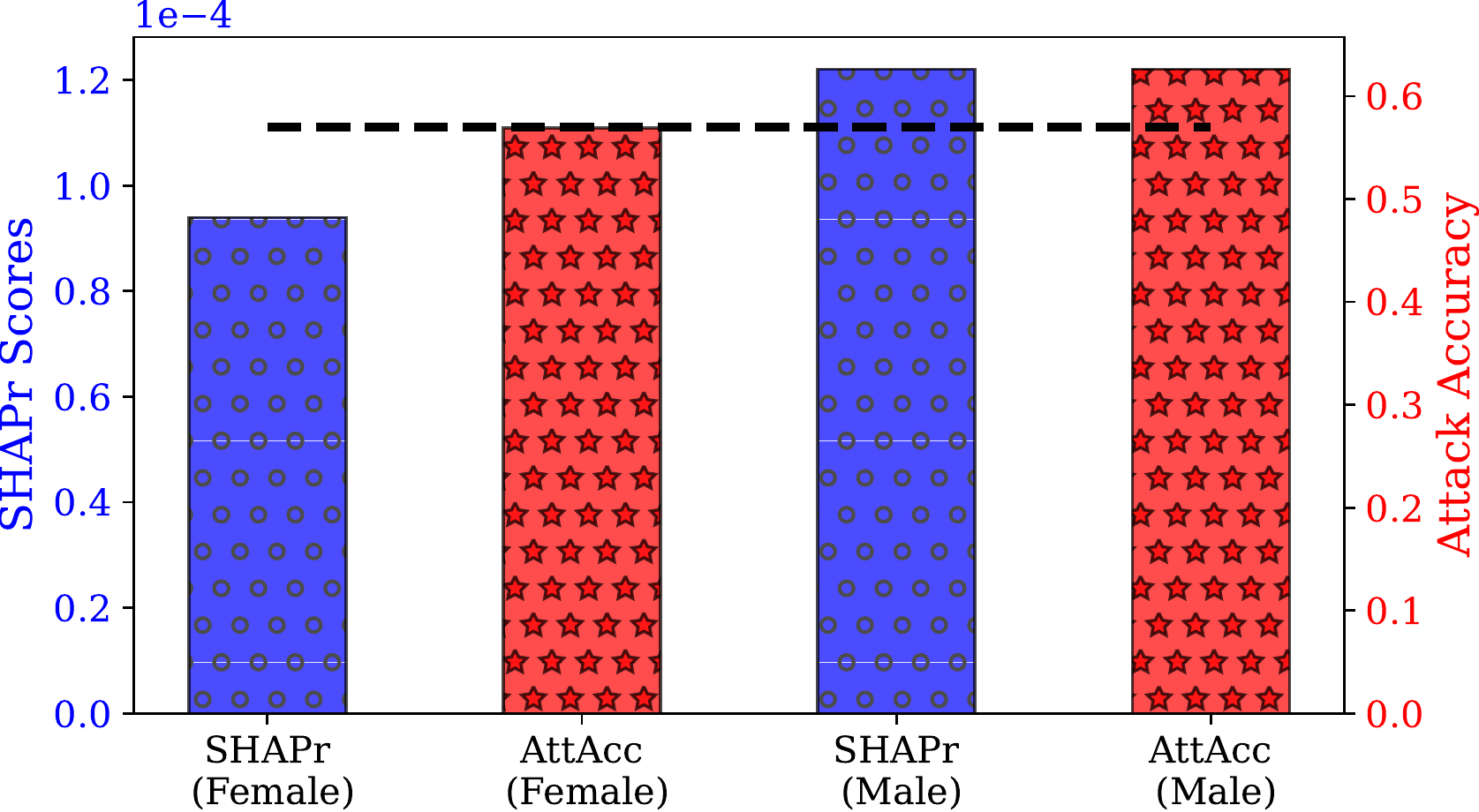}}
    \caption{Different subgroups are vulnerable to \mia{s} to a different extent. \colorbox{blue!45}{blue} bars indicate \method scores for different groups (read values from left axis). \colorbox{red!45}{red} bars indicate \iment accuracy for different groups (red values from the right axis). Random attack accuracy is indicated by the ``Black dashed line''.}
    \label{fig:fairnessfig}
\end{figure}

We used only three datasets that have sensitive attributes: CENSUS, CREDIT, and MEPS.
CENSUS has two sensitive attributes, gender and race, while CREDIT and MEPS have gender.
For gender, the majority class is ``Male'' and the minority class is ``Female''.
For race, ``White'' is the majority class and ``Black'' is the minority class.
We computed the ground truth \iment accuracy, separately for each class. 

Figure~\ref{fig:fairnessfig} shows that there is a difference in the ground truth \iment accuracy for different subgroups.
\method can capture this -- the scores are higher for subgroups with higher \iment accuracy.
\method scores are additive (Property~\ref{prop2}), and hence, we can compute the membership privacy risk of subgroups by averaging the scores within each subgroup.
Unlike \method, \song do not follows the trend of \iment for different subgroups (see Appendix~\ref{app:versatility}) and therefore are not suitable for evaluating the privacy risk of subgroups.

\subsubsection{Data Valuation}\label{sec:dataval} 
We briefly discuss the application of \method for data valuation.
We did not carry out separate experiments but refer to the extensive prior literature on the use of Shapley values for data valuation \cite{jia2019knn,jia19shapley,ghorbani19datashapley,jia2021scalability}.

Two relevant properties of Shapley values are additivity (Property~\ref{prop2}) which includes \textit{group rationality}, where the complete utility is distributed among all training data records, and heterogeneity (Property~\ref{prop4}), which indicates \textit{equitable assignment} of model utility to training data records based on their influence.
These make Shapley values useful for data valuation~\cite{jia19shapley,ghorbani19datashapley}. Since \method uses Shapley values, once computed, \method scores can be used directly for data valuation of both individual data records as well as groups of data records.

On the other hand, \song cannot be used for data valuation as described in Appendix~\ref{app:versatility}.

\section{Performance Evaluation of \method}\label{sec:perfeval}

We now evaluate the efficiency of \method (requirement~\ref{req4}) and show that \method scores can be computed in reasonable time.
We ran the evaluation on Intel Core i9-9900K CPU @ 3.60GHz with 65.78GB memory. We use the python function \textit{time()} in the \textit{time} library which returns the time in seconds (UTC) since the epoch start.

\setlength\tabcolsep{2pt}
\begin{table}[!htb]
\centering
\caption{Performance of \method across different datasets averaged over ten runs.}
\footnotesize
\begin{tabular}{ |c|c|c|c| } 
 \hline
 \textbf{Dataset} & \textbf{\# Records} & \textbf{\# Features} & \textbf{Execution Time (s)} \\ 
 \hline
  \multicolumn{4}{|c|}{\textbf{\song Datasets}}\\
  \hline
 \textbf{LOCATION} & 1000 & 446 & 130.77 $\pm$ 3.90 \\ 
 \textbf{PURCHASE} & 19732 & 600 & 3065.58 $\pm$ 19.24  \\ 
 \textbf{TEXAS} & 10000 & 6170 & 5506.79 $\pm$ 17.47 \\ 
  \hline
 \multicolumn{4}{|c|}{\textbf{Additional Datasets}}\\
 \hline
 \textbf{MNIST} & 60000 & 784 & 2747.41 $\pm$ 22.65 \\
 \textbf{FMNIST} & 60000 & 784 & 3425.90 $\pm$ 34.03 \\
 \textbf{USPS} & 3000 & 256 & 238.67 $\pm$ 1.74\\ 
 \textbf{FLOWER} & 1500 & 2048 & 174.27 $\pm$ 11.74 \\ 
 \textbf{MEPS} & 7500 & 42 & 732.43 $\pm$ 4.95 \\ 
 \textbf{CREDIT} & 15000 & 24 & 1852.66 $\pm$ 30.92 \\ 
 \textbf{CENSUS} & 24000 & 103 & 3718.26 $\pm$ 18.25 \\ 
 \hline
\end{tabular}
\label{tab:performance_eval}
\end{table} 

Table~\ref{tab:performance_eval} shows the average execution time for computing \method scores across datasets of different sizes over ten runs.
Computation time for \method scores ranges from $\approx2$ mins for LOCATION dataset to $\approx91$ mins for TEXAS.
Since the scores are computed once and designed for \mb with substantial computational resources (e.g., GPUs), these execution times are reasonable.

We first compare \method{'s} efficiency with the closely related \loo based metric proposed by Long et al.~\cite{long2017measuring}.
Long et al.'s na\"{i}ve \loo scores require training $|\dtrain|$ additional models~\cite{long2017measuring} (compared to training a single model for \method).
For the benchmark, we used a subset of the LOCATION dataset with $100$ training data record.
\method is $\approx100\times$ faster than a na\"{i}ve \loo based approach: $3640.21\pm244.08$s (\loo) vs. $34.65\pm1.74$s (\method).
For larger datasets \loo would take unreasonably long time to finish.

Compared to \song, we acknowledge that \song is about $\approx2\times$ faster than \method.
We report the results for \song for a few datasets: LOCATION ($59.78\pm0.28$), FLOWER ($77.56\pm10.01$) and USPS ($104.59\pm6.39$).

Although \song is faster, \method is more effective, especially in terms of its future-proofness (Section~\ref{sec:miaeval}).
Moreover, \method can be used to evaluate membership privacy risk with respect to sensitive subgroups (Section~\ref{sec:versatility}) where \song does not perform well (Appendix~\ref{app:versatility} and~\ref{app:robustness}).
Hence, \method has many benefits over \song that warrant its use, despite being slower.

\section{Related Work}\label{sec:related}


\noindent\textbf{Estimating influence of training data record.}
Data marketplaces trade training data for \ml models.
They assign monetary value to data by estimating the influence of each training data record to the model utility.
It was shown that the influence can be measured using influence functions~\cite{influencefnc} or by comparing the gradients produced by a record with respect to itself and other records during training (TracIN~\cite{tracin}).
However, these approaches are computationally expensive and do not precisely estimate memorization for membership privacy risk (Section~\ref{sec:influencefnc}). 
Influence functions can be computed using the \loo approach~\cite{feldman2020memorization}.
However, this is also computationally expensive.

\noindent\textbf{Measuring Membership Privacy Risk.} 
Adversary's membership privacy advantage~\cite{yeom,Jayaraman2021RevisitingMI} is another metric for evaluating differential privacy mechanisms.
However, it is an aggregate metric and estimates membership privacy risk across all data records.
Fisher Information, originally proposed to compute the influence of the attributes towards the model utility (for attribute inference attacks), was suggested as a metric to estimate membership privacy risk~\cite{hannun2021measuring}.
However, this is limited to linear models with convex loss which does not apply to the neural networks we consider.
Furthermore, computing Fisher information is computationally expensive for large models as it requires inverting a Hessian.
Finally, maximal information leakage~\cite{saeidian2020quantifying} was proposed as a membership privacy risk metric which is an upper bound on the privacy risk for the PATE differential privacy framework~\cite{pate}.
However, this information leakage metric is not designed for individual training data records.

Recent and concurrent works by Yu et al.~\cite{yu2022per} and Carlini et al.~\cite{carlini2022privacy} explore per-record privacy risk metrics. Yu et al.~\cite{yu2022per} propose per-instance accounting of privacy for releasing models trained with DP-SGD.
Carlini et al.~\cite{carlini2022privacy} also note that different records are memorized to different extents and explore the reasons behind this phenomenon using scores generated from \ilira.
However, their approach is computationally expensive as it requires training a large number of ML models.
In contrast, \method is an efficient approach for quantifying memorization in any \ml model.

\section{Discussion}\label{sec:discussions}

We first discuss the viability of influence functions as an alternative membership privacy risk metric (Section~\ref{sec:influencefnc}), followed by a note on \method scores for backdoors (Section~\ref{disc:backdoors}).

\subsection{Comparison with Influence Functions}\label{sec:influencefnc}

We discuss alternatives to \method, specifically \textit{influence functions}, and their limitations as privacy risk metrics 
Influence functions~\cite{influencefnc,tracin} were proposed for explaining model predictions.
Since these are independent of specific \mia{s} (satisfying attack-agnostic requirement~\ref{req1} similar to \method), they could potentially be used to design an alternative, interpretable (satisfy Property~\ref{prop1} similar to \method) metric for measuring membership privacy risk. We now explore the viability of such designs. 

We implemented Koh et al.'s influence function~\cite{influencefnc} (referred to as \koh) and TracIN~\cite{tracin}.
To estimate the scores of training data records across the entire test dataset, we averaged the values across all the testing data records for each training data record as suggested by the authors~\cite{influencefnc,tracin}.
For evaluation, we compute recall by thresholding \koh and TracIN scores at zero.
We then compare them with the \mia success, using \iment as the ground truth.

\setlength\tabcolsep{2pt}
\begin{table}[!htb]
\caption{Effectiveness of blackbox influence functions (\koh~\cite{influencefnc}) and TracIN~\cite{tracin} as a metric for membership privacy risk scores with respect to \iment. Comparing recall to \method for \iment from Table~\ref{tab:recall}: \colorbox{orange!25}{orange} indicates comparable results,
\colorbox{red!25}{red} indicates that \koh/TracIN result is worse than \method. \colorbox{green!25}{green} indicates that \koh/TracIn is better than \method. Computations which took unreasonably long time were omitted, indicated by ``-''.}
\footnotesize
\begin{center}
\begin{tabular}{ |c|c|c|c| } 
 \hline
\multirow{2}{*}{\textbf{Dataset}} &
\multicolumn{1}{c|}{\textbf{\method}} & \multicolumn{1}{c|}{\textbf{\koh}~\cite{influencefnc}} &  \multicolumn{1}{c|}{\textbf{TracIN}~\cite{tracin}}\\
& \textbf{Recall}  & \textbf{Recall}\\
 \hline
   \multicolumn{4}{|c|}{\textbf{\song Datasets}}\\
  \hline
 \textbf{LOCATION} & 0.87 $\pm$ 0.01 & \cellcolor{red!25}0.48 $\pm$ 0.01 & \cellcolor{red!25}0.20 $\pm$ 0.00 \\ 
 \textbf{PURCHASE} & 0.81 $\pm$ 0.01 & \cellcolor{red!25}0.51 $\pm$ 0.01  & - \\ 
 \textbf{TEXAS} & 0.73 $\pm$ 0.03  & \cellcolor{red!25}0.51 $\pm$ 0.03 & - \\ 
 \hline    
  \multicolumn{4}{|c|}{\textbf{Additional Datasets}}\\
 \hline
 \textbf{MNIST} & 0.94 $\pm$ 0.00 & \cellcolor{red!25}0.30 $\pm$ 0.18 & - \\ 
 \textbf{FMNIST} & 0.89 $\pm$ 0.03 & \cellcolor{red!25}0.49 $\pm$ 0.10 & - \\ 
 \textbf{USPS} & 0.76 $\pm$ 0.07 & \cellcolor{red!25}0.33 $\pm$ 0.10 & \cellcolor{red!25}0.42 $\pm$ 0.03\\ 
 \textbf{FLOWER} & 0.94 $\pm$ 0.01  & \cellcolor{red!25}0.51 $\pm$ 0.07 & \cellcolor{red!25}0.46 $\pm$ 0.10 \\ 
 \textbf{MEPS} & 0.91 $\pm$ 0.01 & \cellcolor{red!25}0.62 $\pm$ 0.05 & \cellcolor{red!25}0.85 $\pm$ 0.00 \\ 
 \textbf{CREDIT} & 0.92 $\pm$ 0.02  & \cellcolor{red!25}0.79 $\pm$ 0.03 & - \\ 
 \textbf{CENSUS} & 0.87 $\pm$ 0.02 & \cellcolor{red!25}0.72 $\pm$ 0.12 & - \\ 
 \hline
\end{tabular}
\end{center}
\label{tab:influence}
\end{table}

We observe that both \koh and TracIN have low recall values compared to the \iment predictions.
\koh is well defined for convex functions but not for large non-convex neural networks~\cite{basu2021influence}.
Hence, influence scores are often erroneous.
Furthermore, compared to \method, their recall is significantly worse and hence not effective (requirement~\ref{req2}).
Additionally, the high computational cost of \koh and TracIN (compared to \method) does not satisfy the efficiency requirement~\ref{req4}: TracIN has complexity of O($N_{models}. |\dtrain| . |\dtest|$) where $N_{models}$ is the number of intermediate models saved during training, required to compute the influence; \koh's complexity is in the order of O($|\dtrain| . |\dtest|$). Hence, our evaluation indicates that both the state-of-the-art influence functions (\koh and TracIN) are not good candidates for  membership privacy risk metrics.

\subsection{Backdoors and \method}\label{disc:backdoors}

A backdoor to a machine learning model is a set of inputs chosen to manipulate decision boundaries of the model.
Backdoors can be used for malicious purposes such as poisoning (e.g.~\cite{chen2017targeted}), or to embed watermarks that allow model owners to claim ownership of their model in case it gets stolen~\cite{szyller2021dawnextended,adi2018turning}.
A backdoor is created by changing the label of several training data records~\cite{szyller2021dawnextended}, by adding artifacts to the training data records themselves (e.g. overlay text or texture to images~\cite{zhang2018protecting}), or by introducing out-of-distribution data~\cite{adi2018turning} to the training data.
A successfully embedded backdoor is memorised during training, along the primary task of the model.
During the verification, a verifier (can either be \mb or a third-party judge where \mb provides the watermark set to the judge) queries the model and expects matching backdoor predictions.

Backdoors have negative influence on model utility as they introduce noise, and make training more difficult~\cite{jia2021scalability}.
Hence, their \method scores are low. 
This has been used as a way for identifying and removing images with watermarks~\cite{jia2021scalability}.

However, memorization of backdoors is required for successful verification.
In other words, backdoors behave differently from other data records in the context of \method: they are, by definition, memorized but unlike other memorized data records, they are likely to have low \method scores.
This is not a concern in our setting because \mb is the entity that computes \method scores.
If a backdoor is inserted intentionally by \mb (e.g., for watermarking), then \mb will know what they are.
If a backdoor was inserted maliciously (e.g., by a training data provider), there is no need to provide any guarantees regarding the \method scores for those records.



\section{Conclusion}\label{sec:conclusions}

Membership privacy risk metrics quantify the susceptibility of training data records to \mia{s}. 
We present the desiderata for designing an effective membership privacy risk metric for quantifying the susceptibility of individual training data records to \mia{s}. 
Our conjecture was that Shapley values computed for individual training data records, by measuring the influence on the model utility, and hence the extent of memorization, can serve as a good membership privacy risk metric while satisfying all the requirements.

We successfully validated our conjecture by presenting \method, a metric based on Shapley values as a membership privacy risk metric. 
By definition of Shapley values, \method is fine-grained and assigns scores for individual training data records without using any specific \mia.
We show that \method indeed serves as an effective membership privacy risk metric to assess susceptibility of different training data records to \mia{s}. \method outperforms prior work (\song) on the state-of-the-art \mia.
Additionally, \method can be used for other applications, e.g., to evaluate defences against \mia{s} and assess the privacy risk of different subgroups. Finally, \method can be computed more efficiently compared to a na\"{i}ive \loo approach.

\section*{Acknowledgement}
This work was supported in part by Intel (in the context of the Private-AI Institute).

{\footnotesize
\bibliographystyle{IEEEtranS}
\bibliography{paper}
}

\appendix

\subsection{Model and Attack Accuracy}\label{app:accuracy}

In Table~\ref{tab:performance}, we summarize the test accuracy for the models built from each dataset, and the corresponding attack accuracy of \iment against each model. We note that the model utility and \iment accuracy on \song datasets is close to reported results in Song and Mittal~\cite{song2020systematic}. We also report the same metrics for the additional datasets.

\setlength\tabcolsep{2pt}
\begin{table}[!htb]
\footnotesize
\caption{Test accuracy of target models, \iment and \ilira accuracy for each dataset averaged over 10 runs.}
\centering
\begin{tabular}{ |c|c|c| } 
\hline
 \textbf{Dataset} & \textbf{Test Accuracy} &  \textbf{\iment} \\
 \hline
  \multicolumn{3}{|c|}{\textbf{\song Datasets}}\\
 \hline
\textbf{LOCATION} & 69.00 & 87.70\\ 
\textbf{PURCHASE} & 84.65 & 64.08\\
\textbf{TEXAS} & 49.92 & 79.43\\
\hline
\multicolumn{3}{|c|}{\textbf{Additional Datasets}} \\
\hline
\textbf{MNIST} & 98.10 & 54.30\\
\textbf{FMNIST} & 89.30 & 57.90\\
\textbf{USPS} & 95.50 & 54.13\\
\textbf{FLOWER} & 89.60 & 68.81\\
\textbf{MEPS} & 84.00 & 61.73\\
\textbf{CREDIT} & 79.90  & 57.18\\
\textbf{CENSUS} & 82.20 & 55.95\\
 \hline
\end{tabular}
\label{tab:performance}
\end{table}

\subsection{Evaluating \song's Versatility}\label{app:versatility}

In Section~\ref{app:versatility}, we argued that \method is versatile. A natural question is whether \song is similarly versatile. We first evaluate whether \song correlates with changes \iment accuracy across different sensitive subgroups. We then discuss whether \song can be used for data valuation.

\noindent\textbf{Privacy Risks over Subgroups.} We compute \iment attack accuracy over different sensitive subgroups and average \song scores for each of the subgroups. We report the results in Table~\ref{tab:song_fairness} which is color-coded: \colorbox{green!25}{green} indicates \song moves in the same direction as the ground truth \iment; and \colorbox{red!25}{red} indicates \song either remains the same or moves in the opposite direction as the ground truth \iment.

\begin{table}[!htb]
\caption{Different subgroups are vulnerable to membership inference attacks (\mia{s}) to a different extent. \colorbox{green!25}{green} indicates \song moves in the same direction as the ground truth \iment. \colorbox{red!25}{red} indicates \song either remains the same or moves in opposite direction as the ground truth \iment.}
\footnotesize
\begin{center}
\begin{tabular}{ |c|c|c|c|c| } 
\hline
\textbf{Dataset} & \multicolumn{2}{|c|}{\song} & \multicolumn{2}{|c|}{\iment}\\
\hline
\multirow{4}{*}{\textbf{CENSUS}} & \textbf{Male} & \textbf{Female} & \textbf{Male} & \textbf{Female} \\
& \cellcolor{red!25}0.52 & \cellcolor{red!25}0.52 & \cellcolor{red!25}56.00 & \cellcolor{red!25}62.50 \\
& \textbf{White} & \textbf{Others} & \textbf{White} & \textbf{Others} \\
& \cellcolor{red!25}0.52 & \cellcolor{red!25}0.52 & \cellcolor{red!25}56.60 & \cellcolor{red!25}60.50 \\
\hline
\multirow{2}{*}{\textbf{CREDIT}} & \textbf{Male} & \textbf{Female} & \textbf{Male} & \textbf{Female} \\
& \cellcolor{red!25}0.52 & \cellcolor{red!25}0.53 & \cellcolor{red!25}56.10 & \cellcolor{red!25}67.00 \\
\hline
\multirow{2}{*}{\textbf{MEPS}} & \textbf{Male} & \textbf{Female} & \textbf{Male} & \textbf{Female} \\
& \cellcolor{red!25}0.57 & \cellcolor{red!25}0.54 & \cellcolor{red!25}56.90 & \cellcolor{red!25}62.60 \\
\hline
\end{tabular}
\end{center}
\label{tab:song_fairness}
\end{table}

We see that \song does not match the trend of \iment accuracy as seen in Table~\ref{tab:song_fairness}. \song either remains the same or moves in opposite direction as the ground truth which is indicated in \colorbox{red!25}{red}.
Table~\ref{tab:song_fairness} shows that \song is ineffective to estimate disparity of membership privacy risk across different sensitive subgroups.


We note that the average scores are close to 0.5 because majority of the data records have \song scores of 0.5 (due to a lack of heterogeneity property \ref{prop4} as seen in Figure~\ref{fig:attplots1} and~\ref{fig:attplots2} in Appendix~\ref{app:alldistributions}).
Additionally, \song do not satisfy additivity property (Property~\ref{prop2}) as there is no semantically meaningful notion of adding or averaging probability scores.
We conjecture that the lack of both heterogeneity and additivity properties make \song makes ineffective at this task.

\noindent\textbf{Data Valuation.} \song was not designed to be additive~\ref{prop2} and hence cannot guarantee group rationality of scores among training data records.
\song are not heterogeneous (Property~\ref{prop4}) either which does guarantee equitable assignment of privacy risk scores (as shown in Appendix~\ref{app:alldistributions}, Figure~\ref{fig:attplots1} and~\ref{fig:attplots2}).
We show the lack of heterogeneity in the Appendix~\ref{app:alldistributions}, visualizing the distribution of \song (Figure~\ref{fig:attplots1} and~\ref{fig:attplots2}).
Given the lack of these properties (heterogeneity, additivity, group rationality, and equitable assignment), we argue that \song is unlikely to be applicable for data valuation.

\subsection{Distribution of \method and \song}\label{app:alldistributions}

\begin{figure*}
    \centering

    \subfigure[MNIST (\method)]{\includegraphics[width=0.49\textwidth]{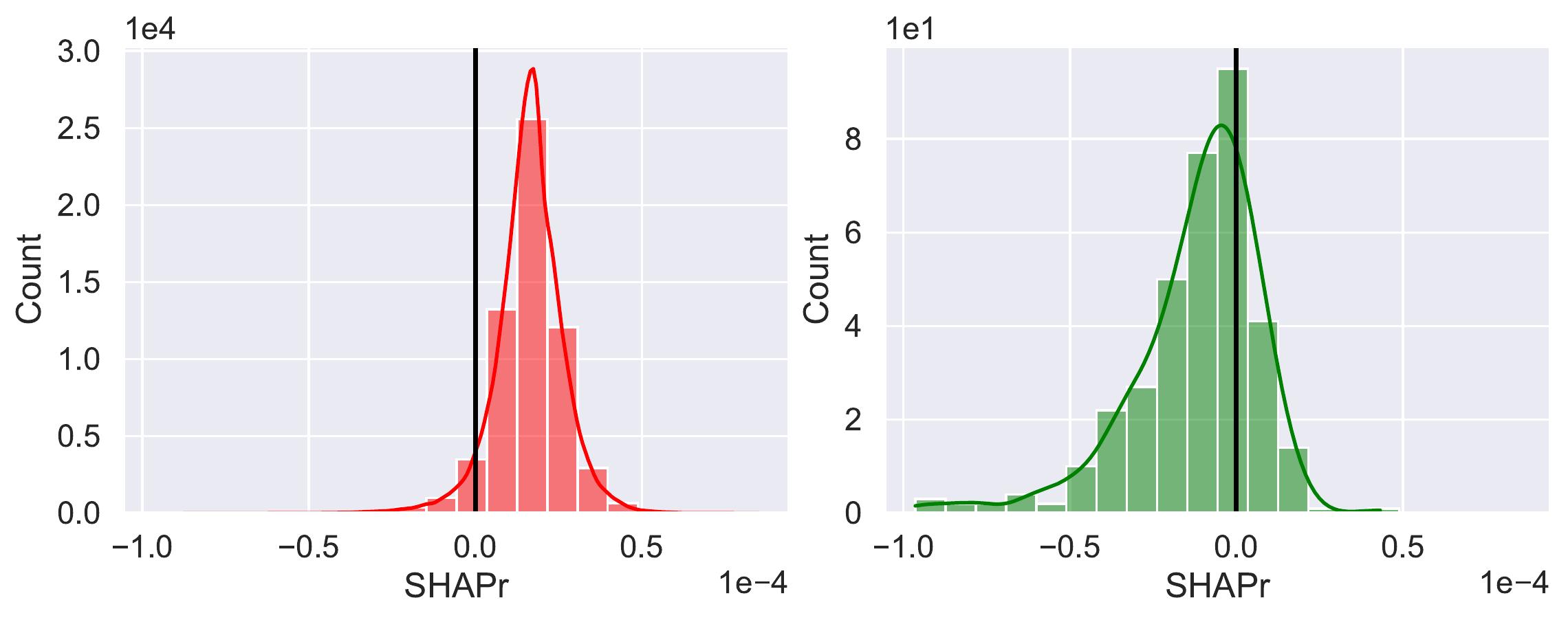}}
    \subfigure[MNIST (\song)]{\includegraphics[width=0.49\textwidth]{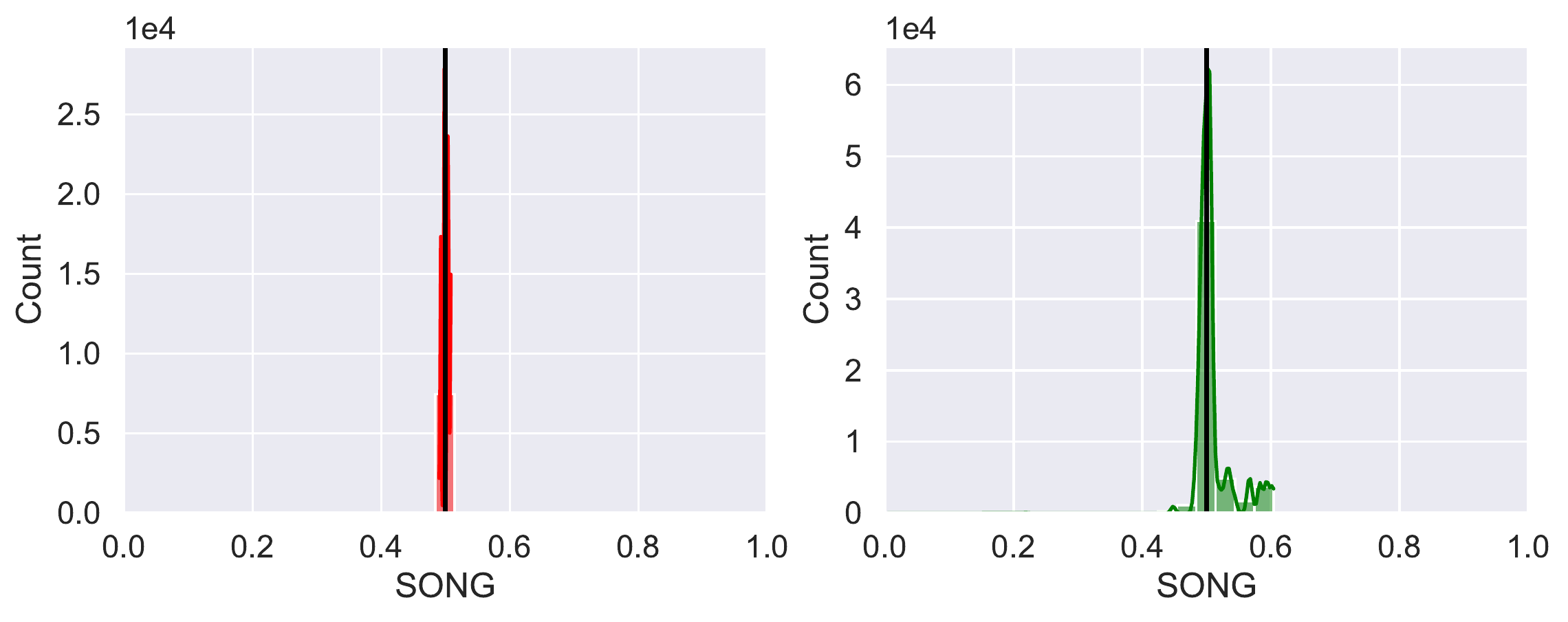}}

    \subfigure[FMNIST (\method)]{\includegraphics[width=0.49\textwidth]{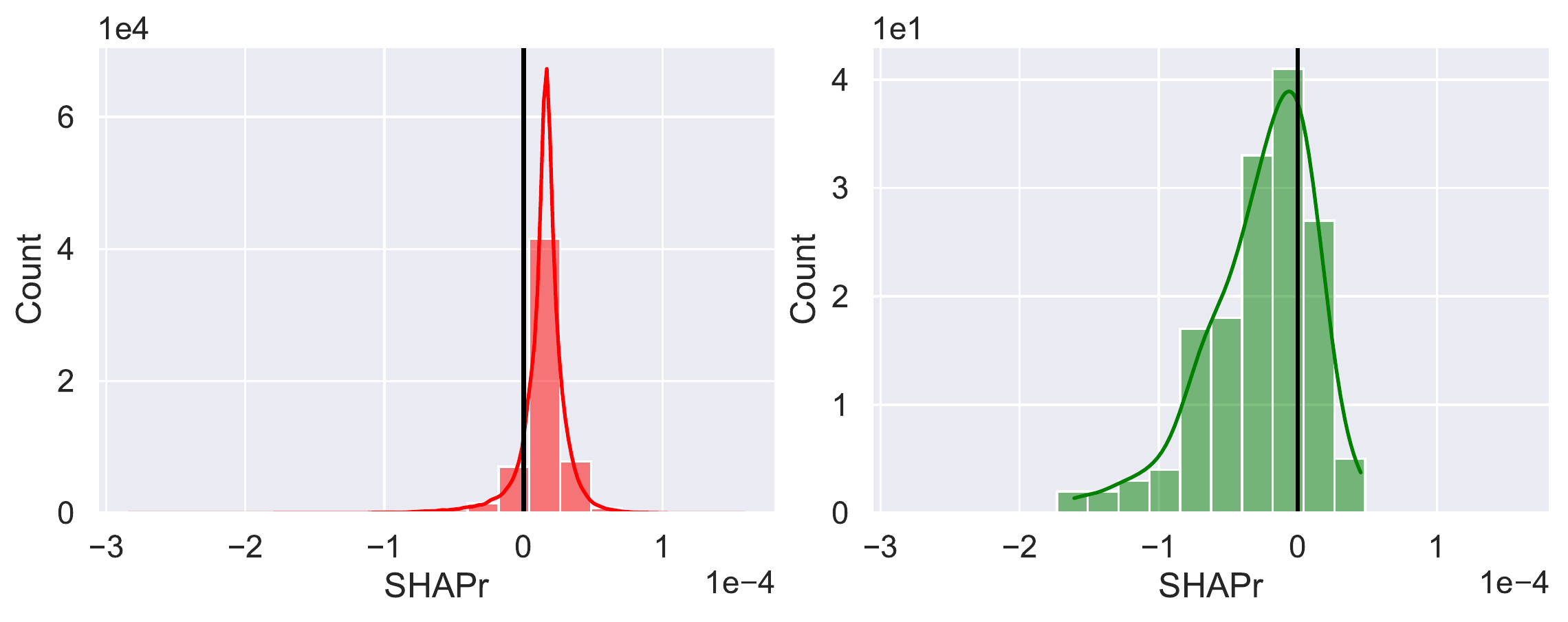}}
    \subfigure[FMNIST (\song)]{\includegraphics[width=0.49\textwidth]{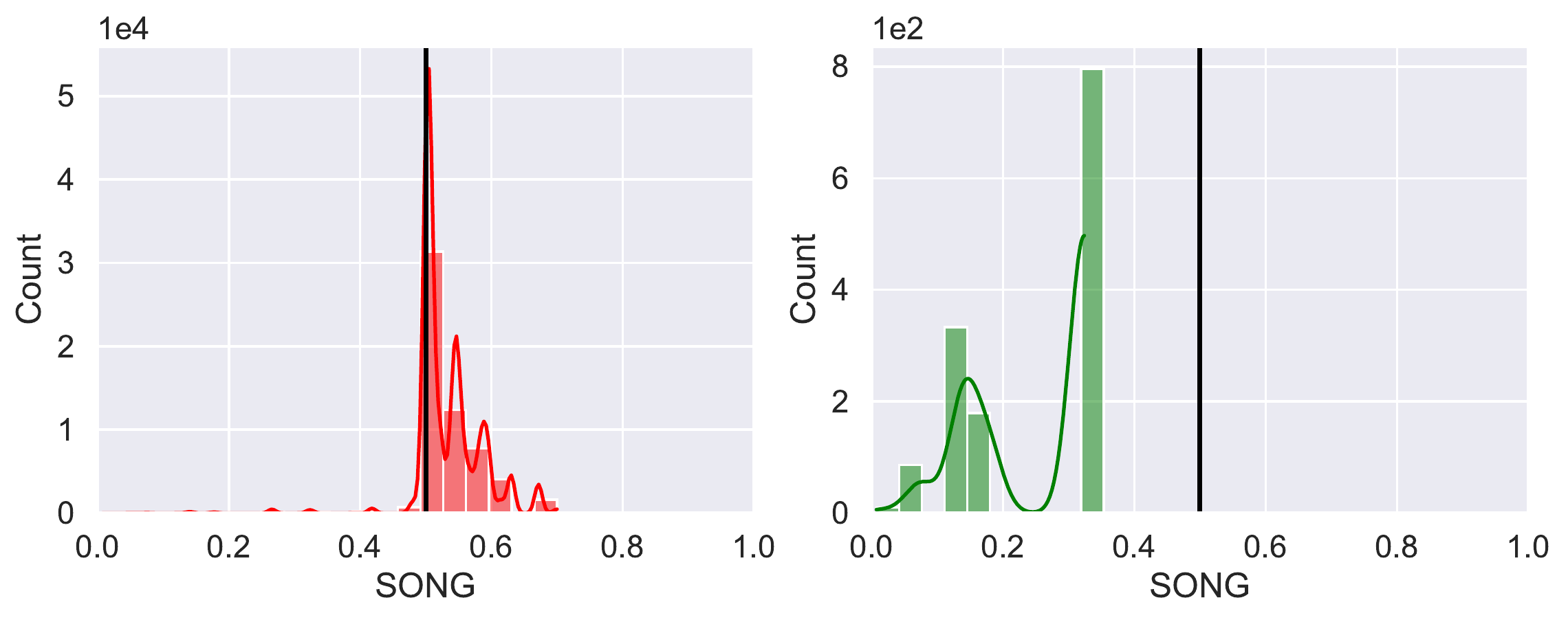}}
  
    \subfigure[CREDIT (\method)]{\includegraphics[width=0.49\textwidth]{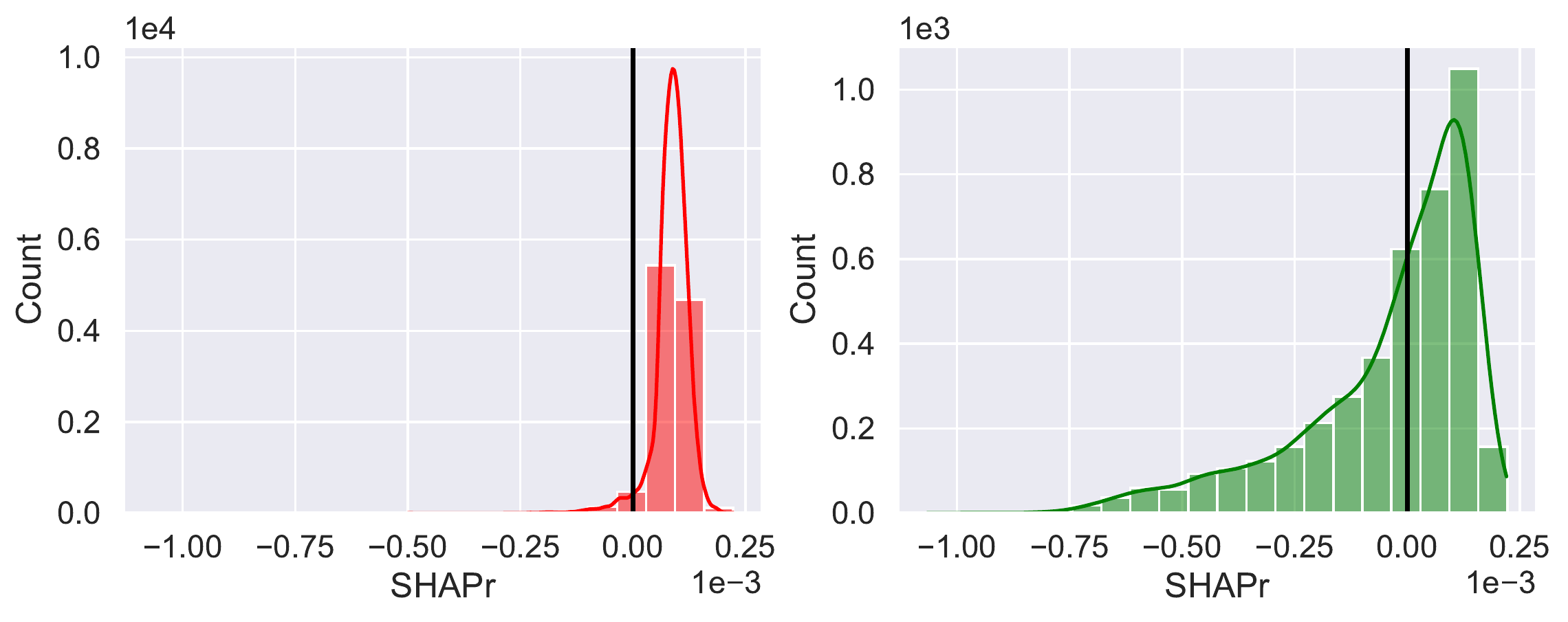}}
    \subfigure[CREDIT (\song)]{\includegraphics[width=0.49\textwidth]{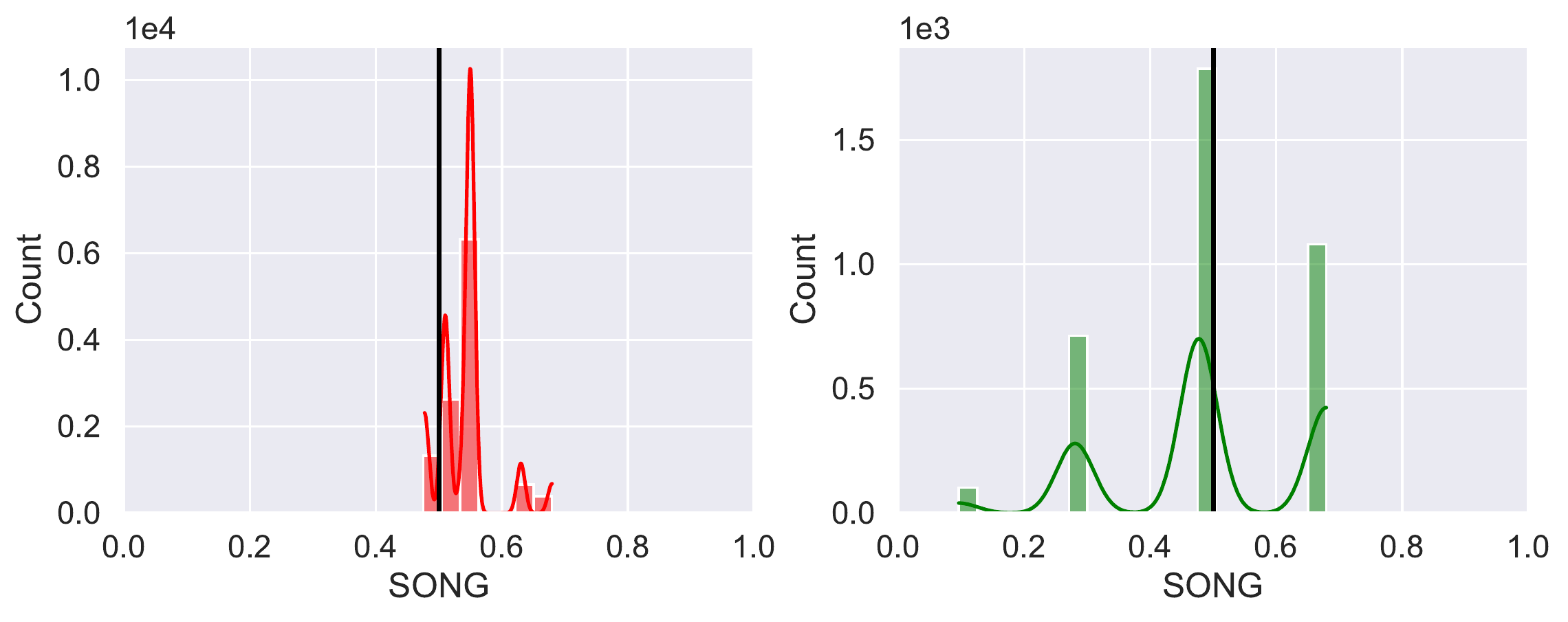}}

    \subfigure[CENSUS (\method)]{\includegraphics[width=0.49\textwidth]{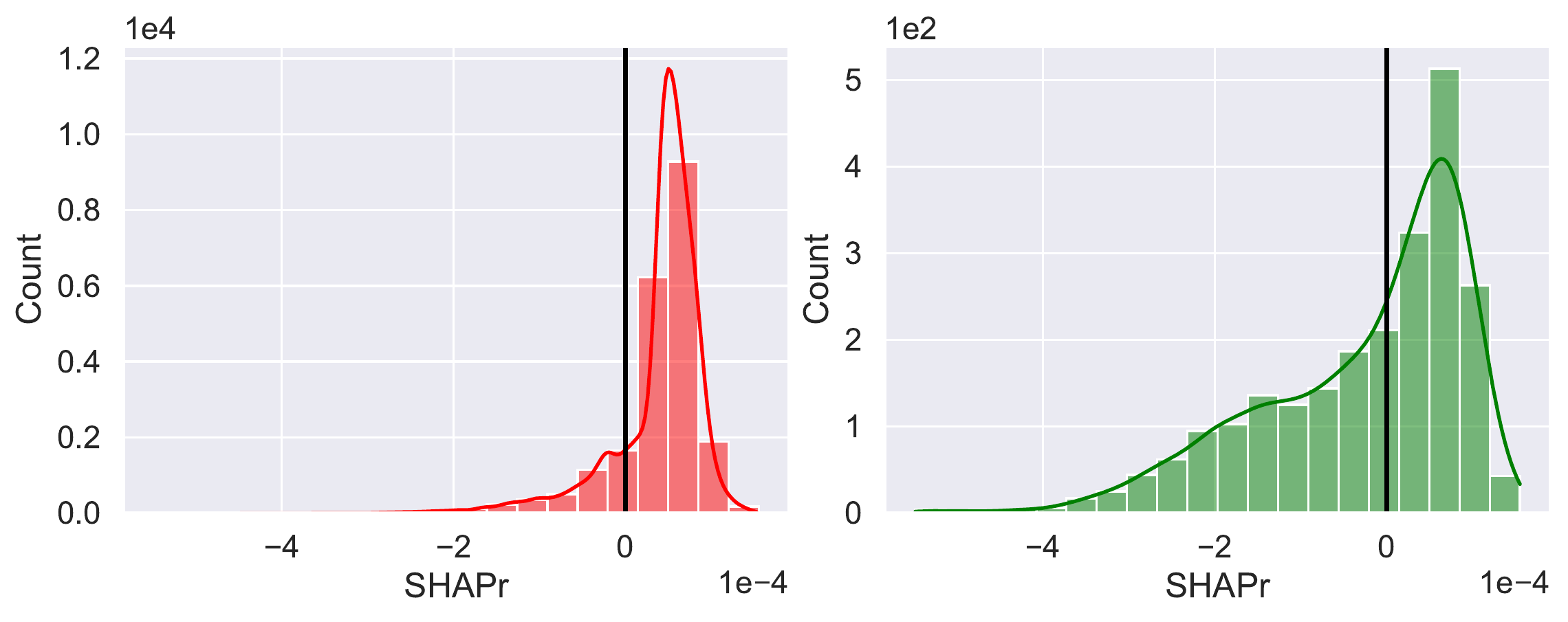}}
    \subfigure[CENSUS (\song)]{\includegraphics[width=0.49\textwidth]{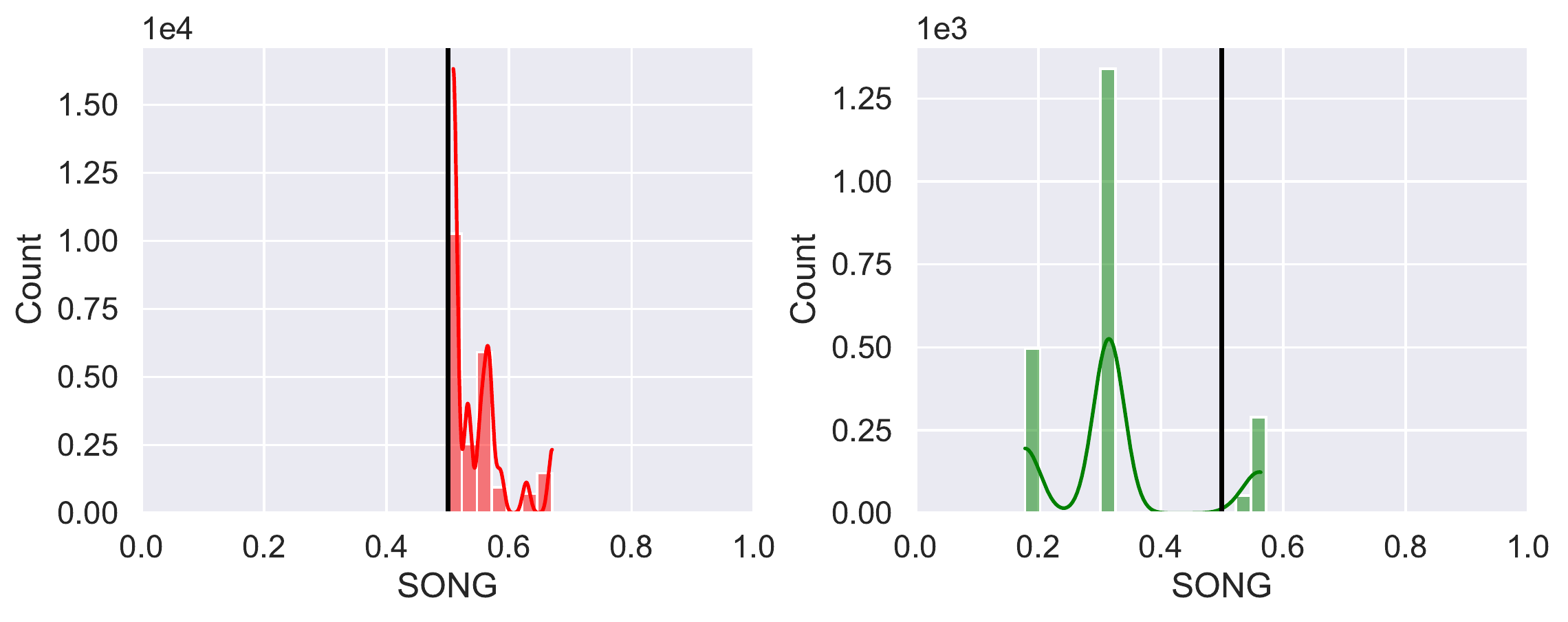}}

    \subfigure[USPS (\method)]{\includegraphics[width=0.49\textwidth]{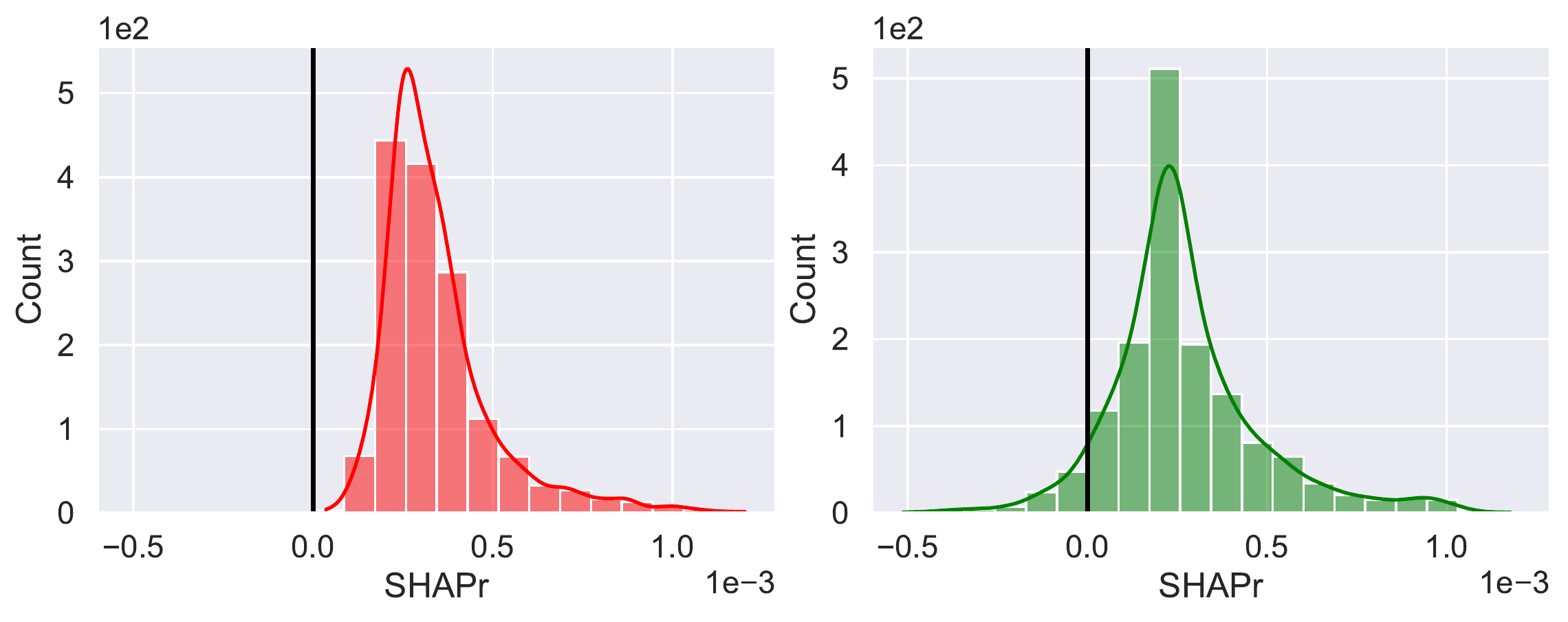}}
    \subfigure[USPS (\song)]{\includegraphics[width=0.49\textwidth]{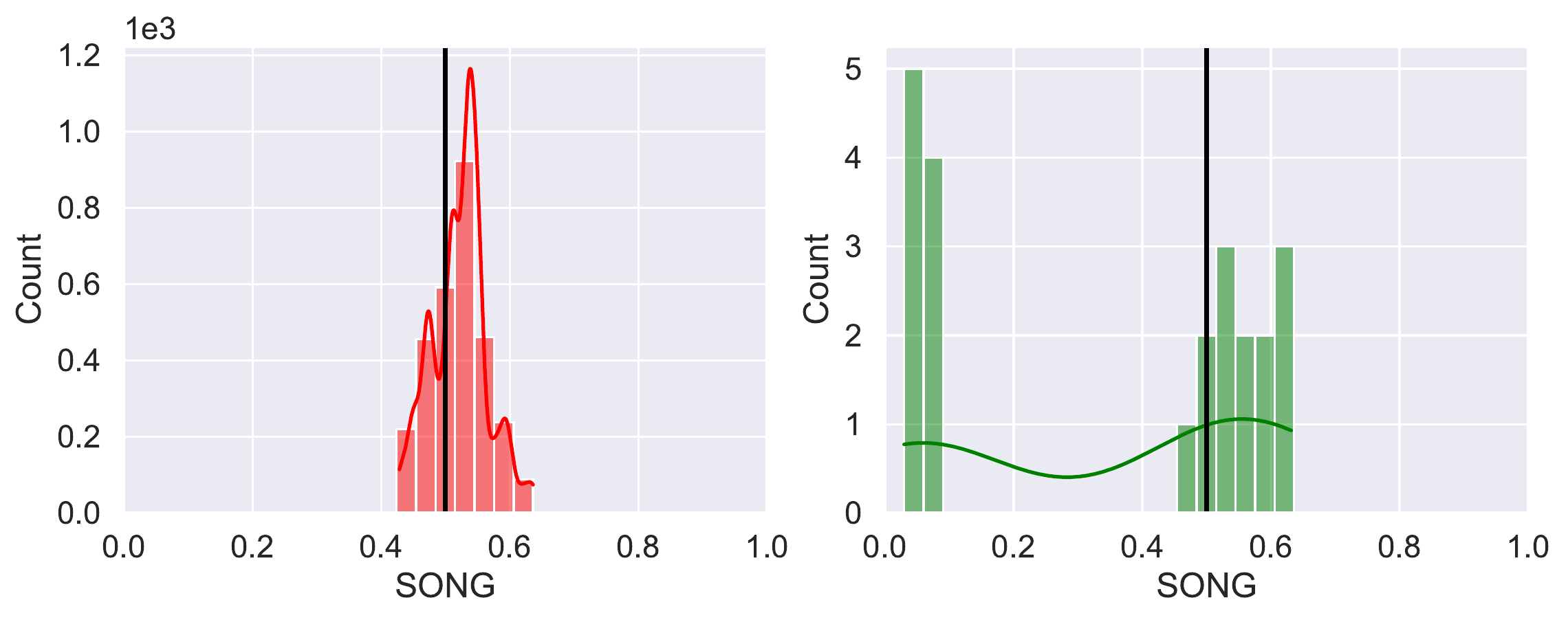}}
    \caption{Distributions of \method scores colored based on \iment; semantic threshold at $0$ for \method and $0.5$ for \song(solid line).}
    \label{fig:attplots1}
\end{figure*}

\begin{figure*}
    \centering
    \subfigure[TEXAS (\method) ]{\includegraphics[width=0.49\textwidth]{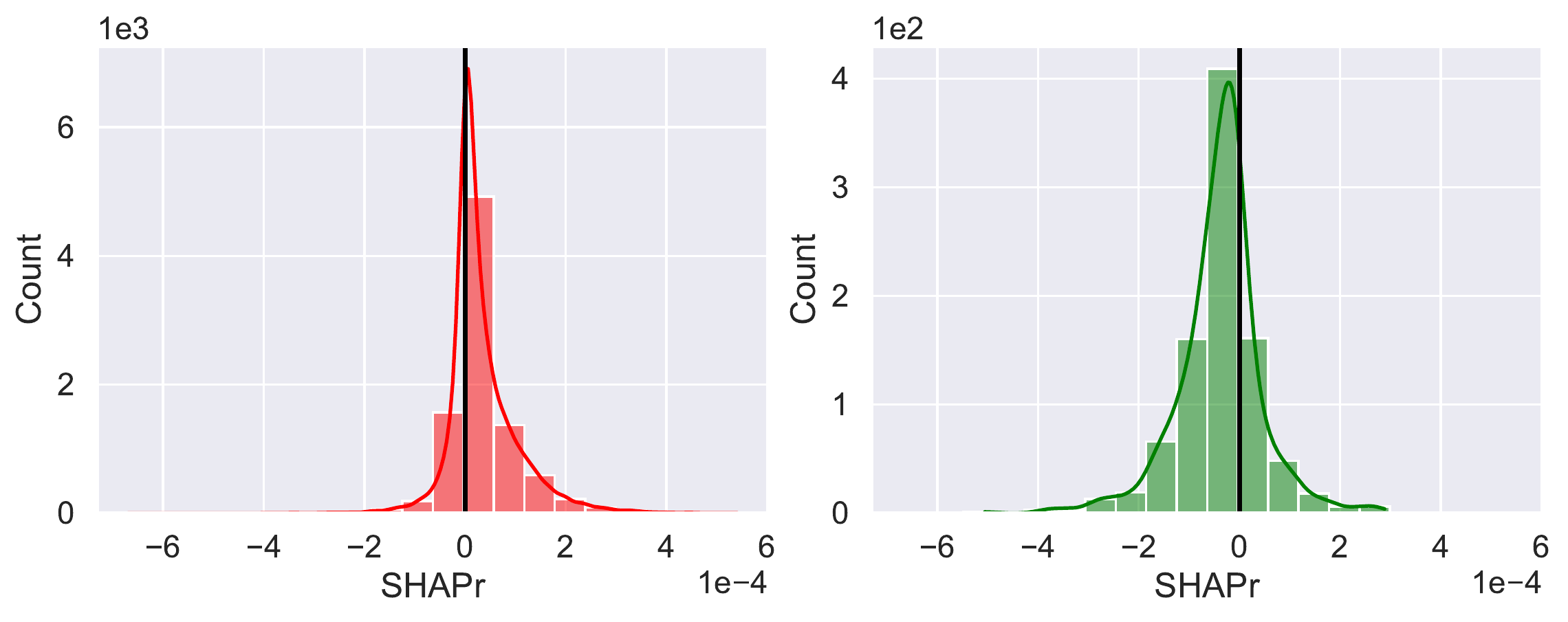}}
    \subfigure[TEXAS (\song) ]{\includegraphics[width=0.49\textwidth]{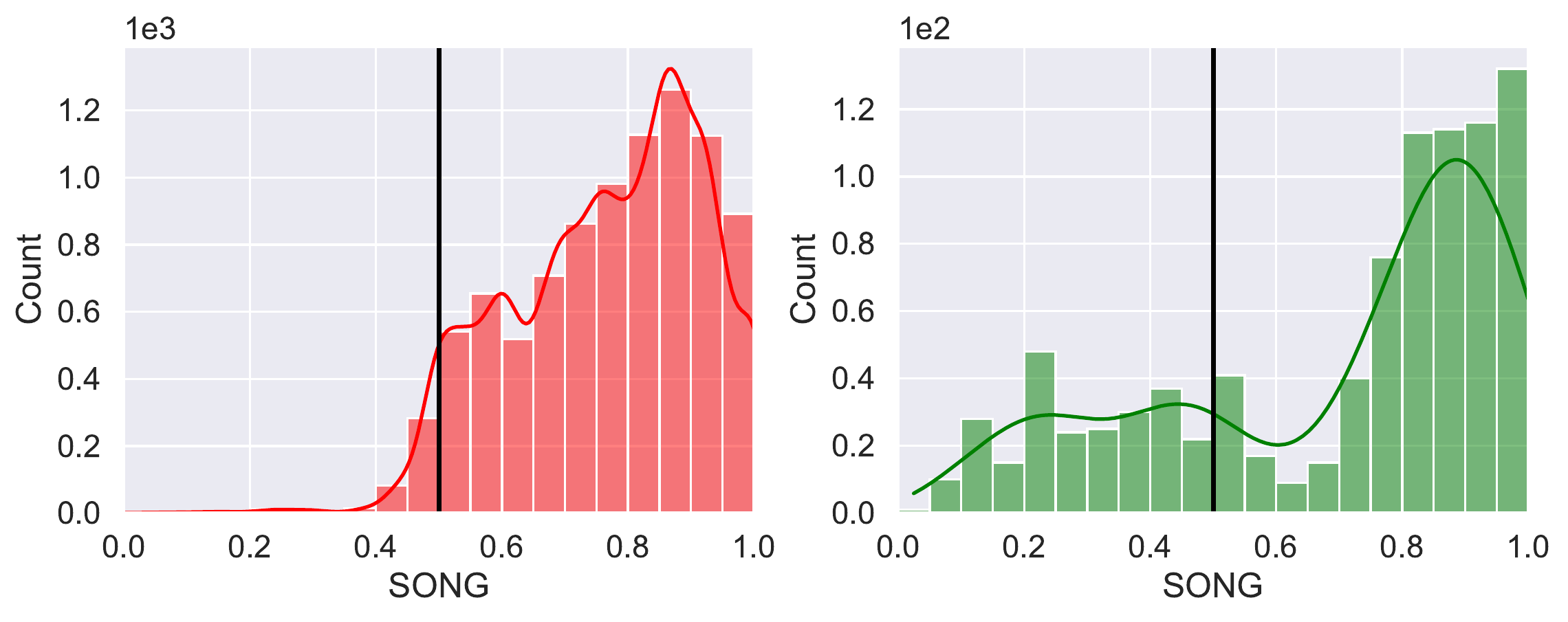}}
    
    \subfigure[MEPS (\method)]{\includegraphics[width=0.49\textwidth]{figures/shapr_CENSUS.pdf}}
    \subfigure[MEPS (\song)]{\includegraphics[width=0.49\textwidth]{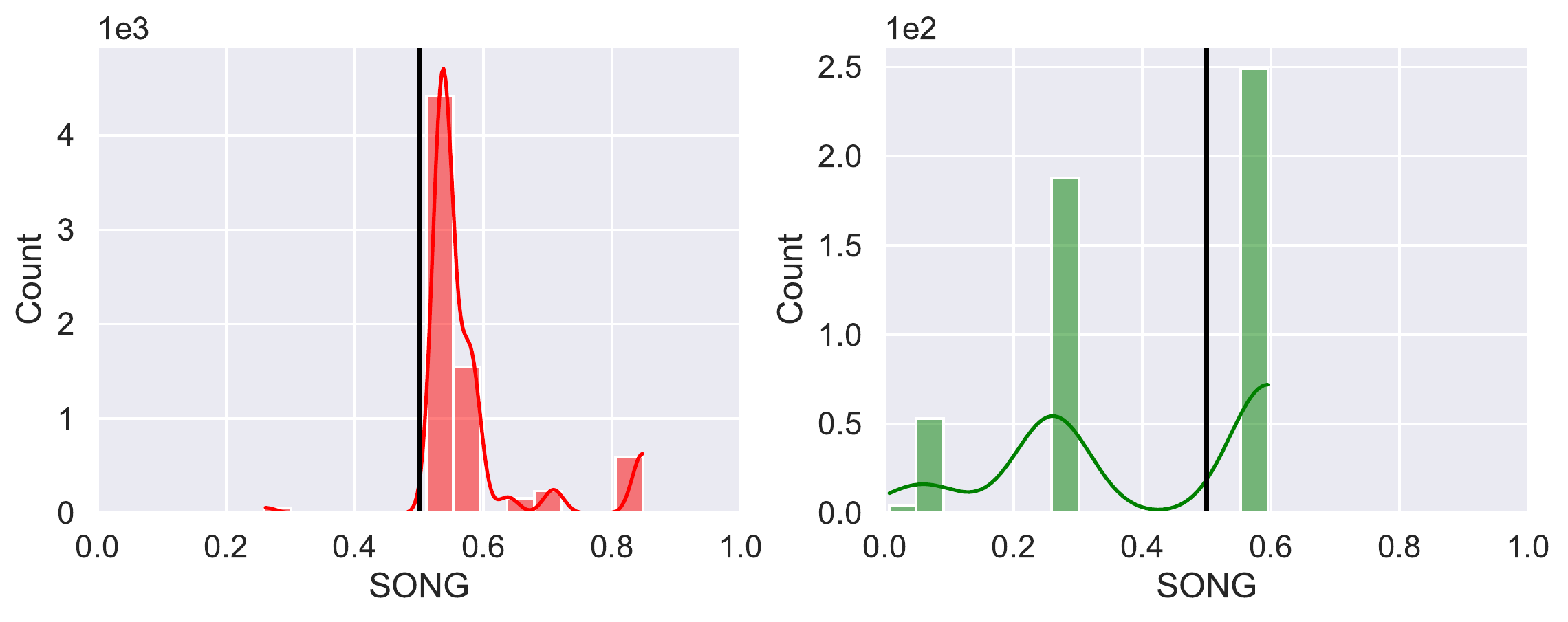}}

    \subfigure[FLOWER (\method) ]{\includegraphics[width=0.49\textwidth]{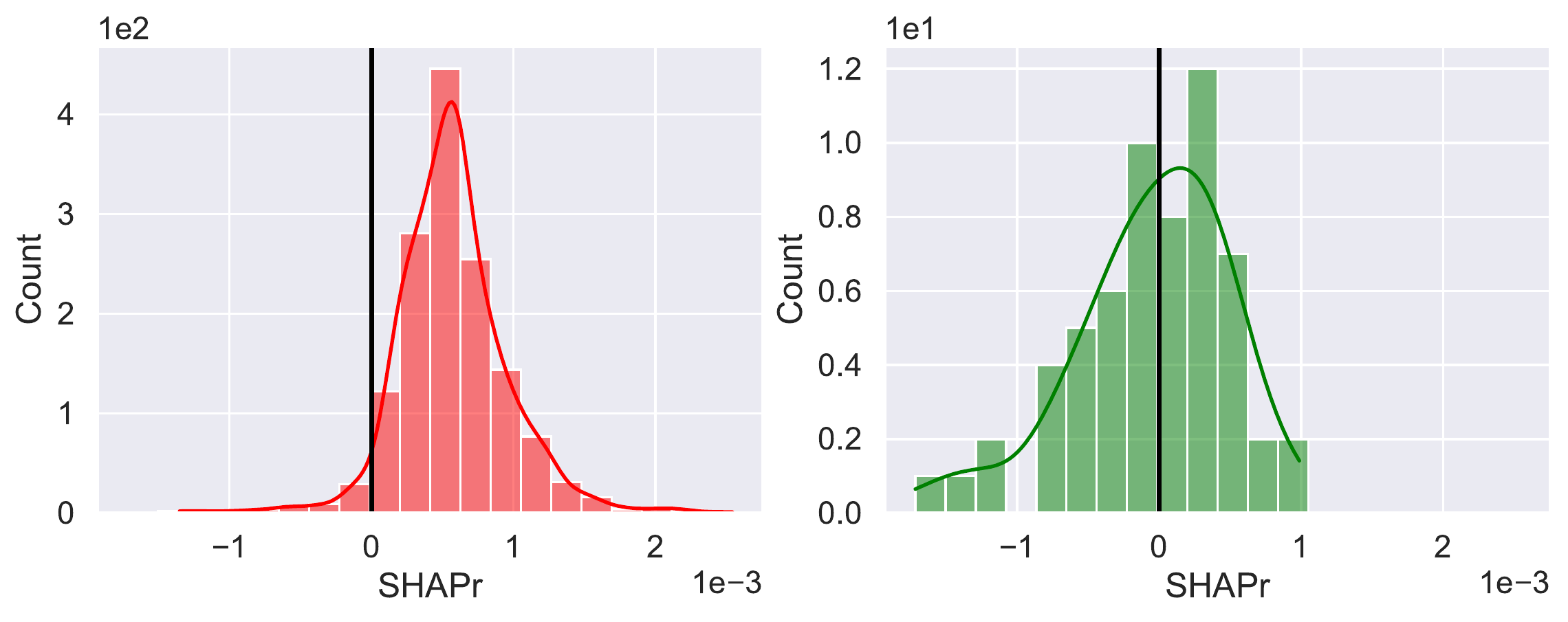}}
    \subfigure[FLOWER (\song) ]{\includegraphics[width=0.49\textwidth]{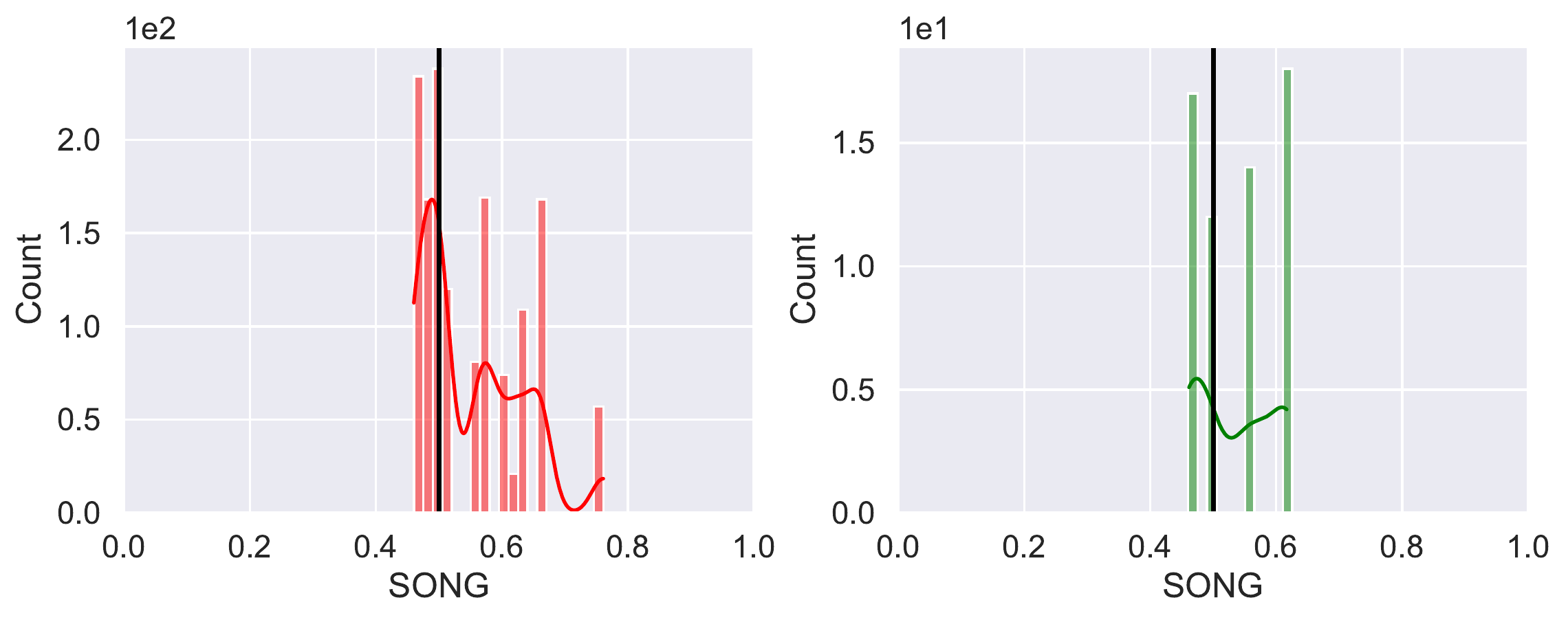}}
    
    \subfigure[LOCATION (\method) ]{\includegraphics[width=0.49\textwidth]{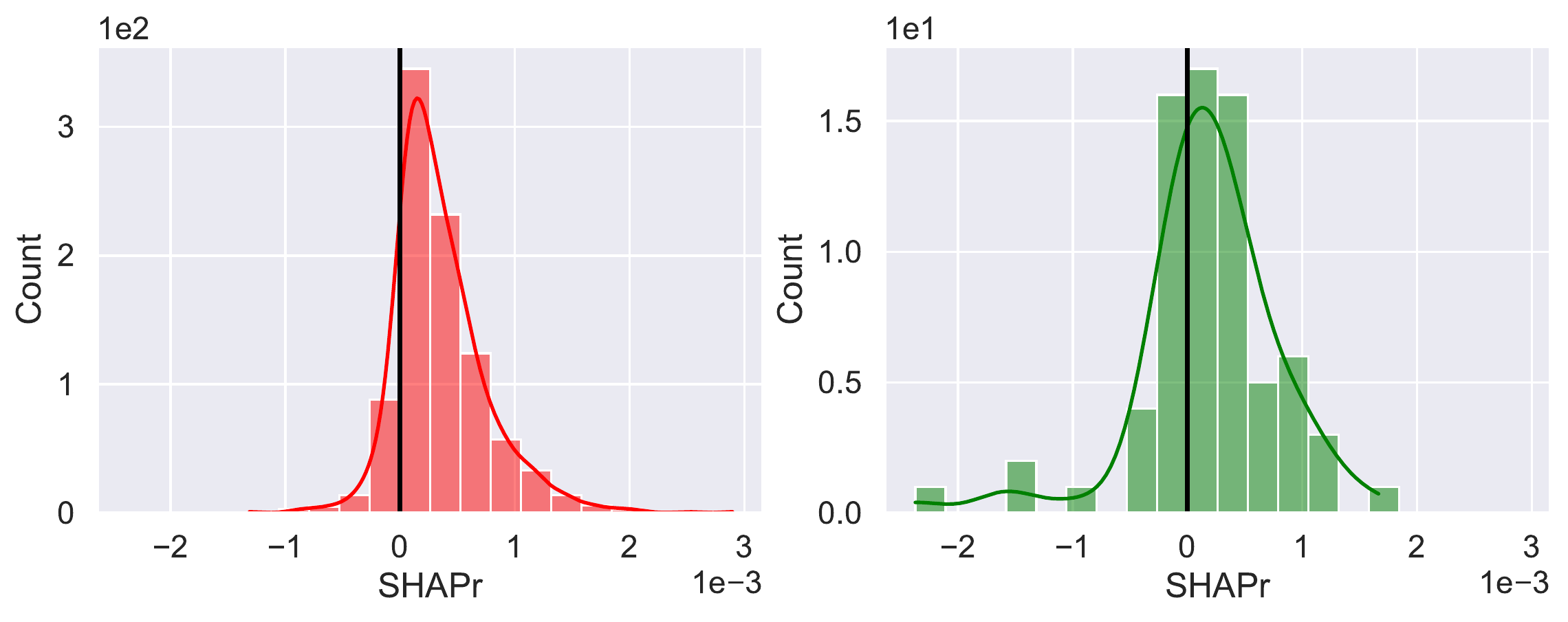}}
    \subfigure[LOCATION (\song) ]{\includegraphics[width=0.49\textwidth]{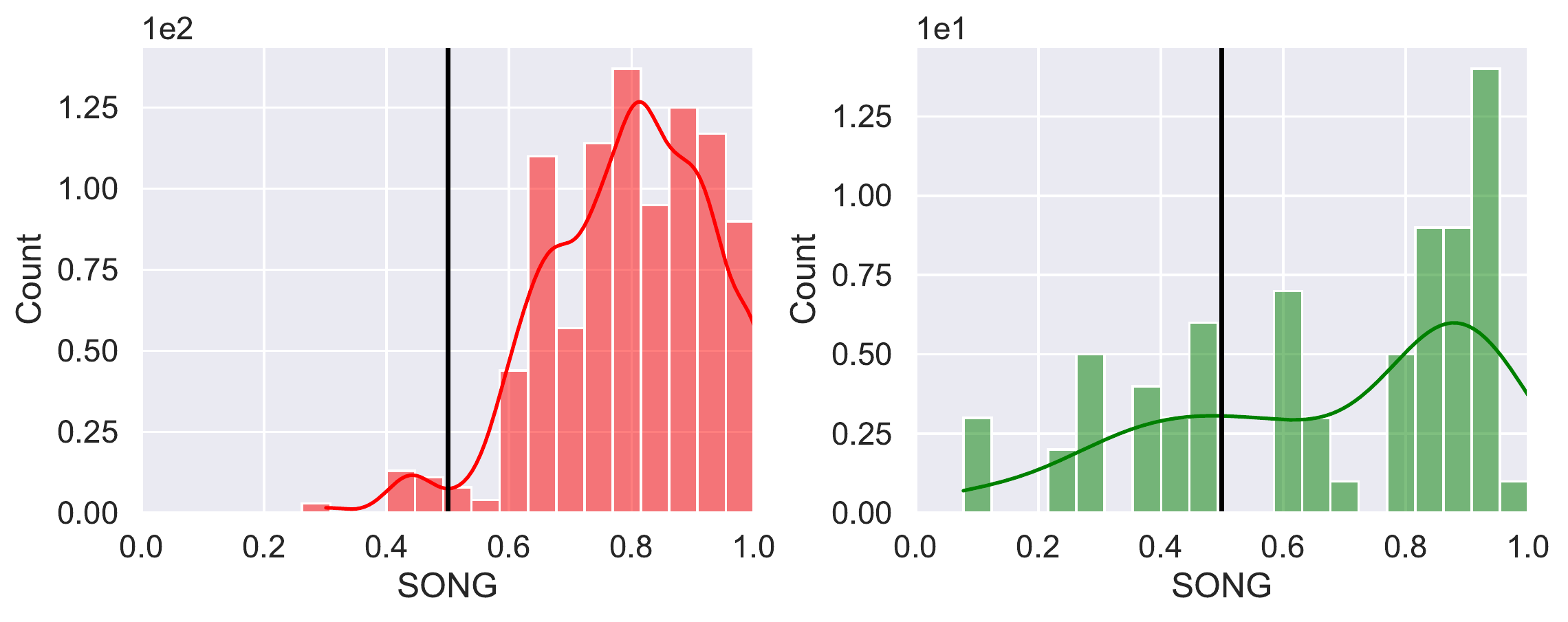}}

    \subfigure[PURCHASE (\method) ]{\includegraphics[width=0.49\textwidth]{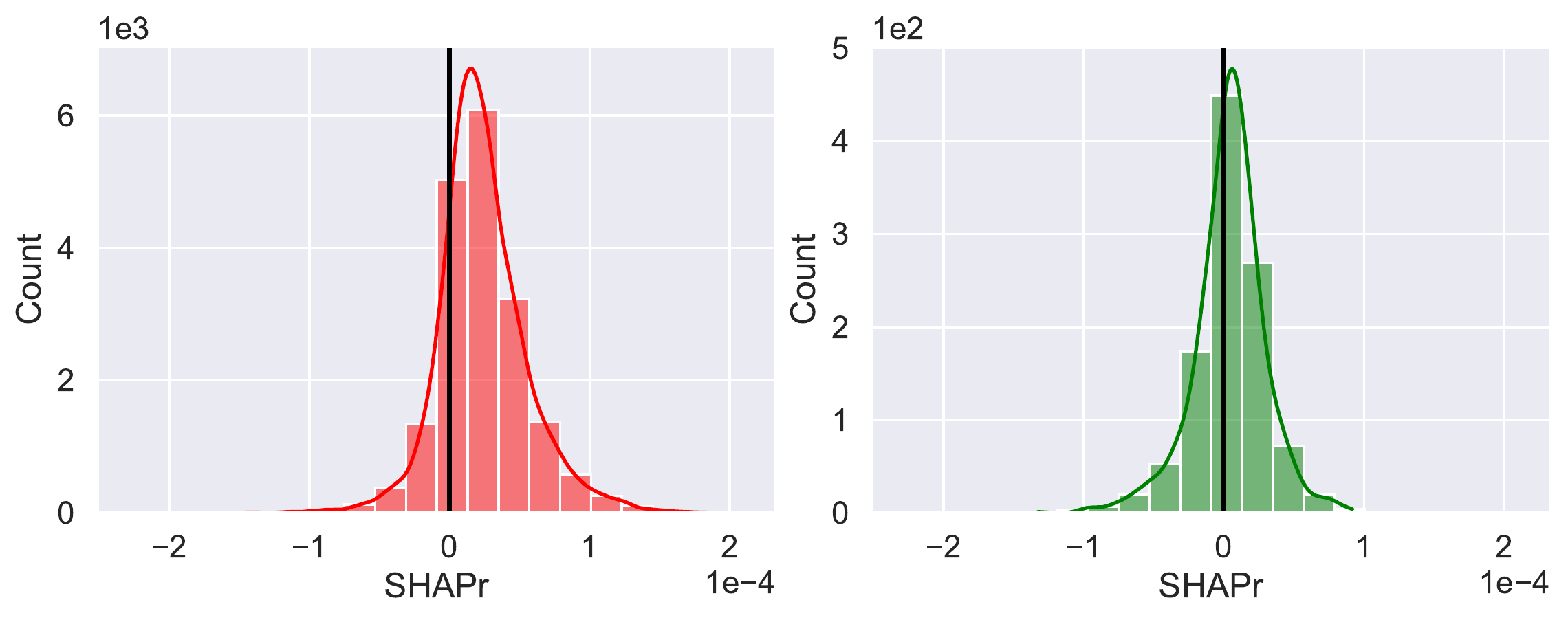}}
    \subfigure[PURCHASE (\song) ]{\includegraphics[width=0.49\textwidth]{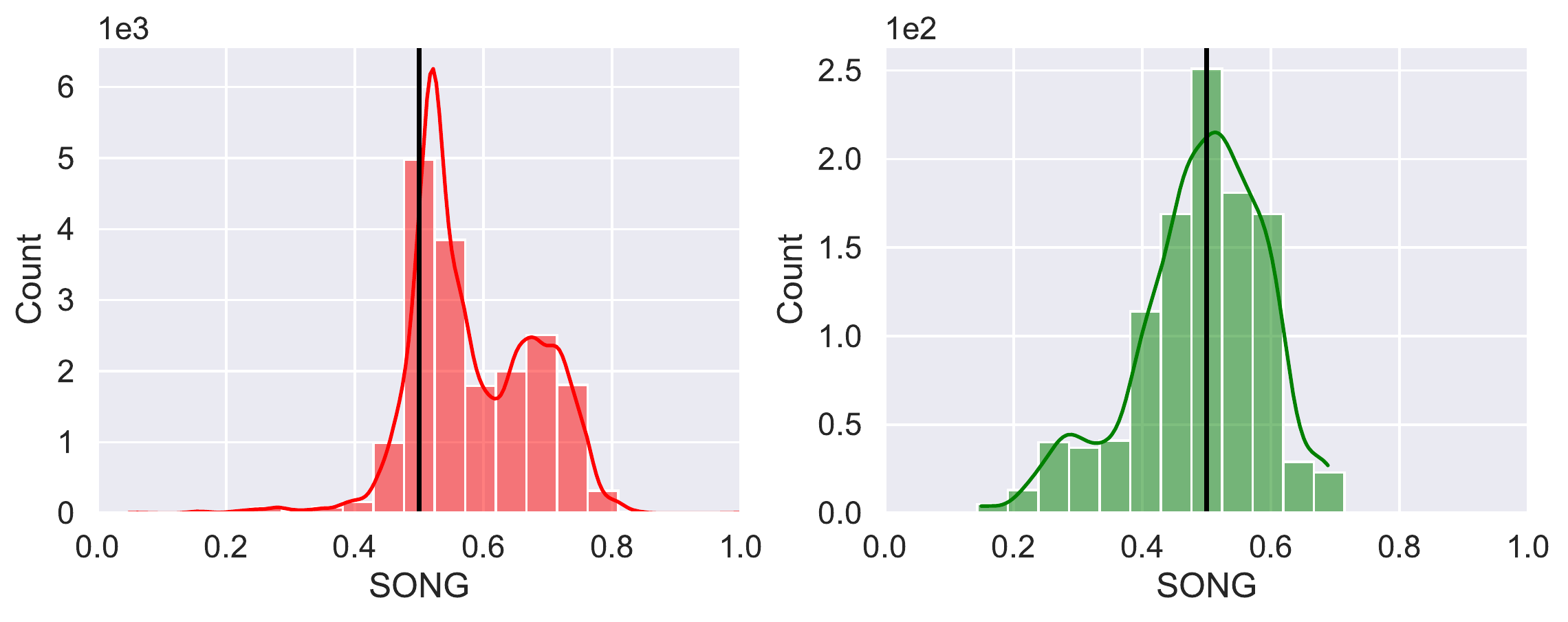}}

    \caption{Distributions of \method scores colored based on \iment; semantic threshold at $0$ for \method and $0.5$ for \song(solid line).}
    \label{fig:attplots2}
\end{figure*}

We visually compare \method with \song by plotting the distribution of \method (in green) and for \song (in red) shown in Figure~\ref{fig:attplots1} and~\ref{fig:attplots2}.
For several datasets, we observe that \song is centered at 0.5 indicating that the membership likelihood for a large number of training data records is inconclusive. Further, we note that the distribution of \song is not evenly distributed, with some values correspond to several records while neighboring values correspond to none. We conjecture that this is due to the fixed prior probabilities and estimating the conditional probabilities using shadow models optimized to give the same output for multiple similar data records.
Compared to \song, \method follows a more even distribution (due to the heterogeneity property \ref{prop4}).

\subsection{Evaluating Effectiveness for Adding Noise}\label{app:robustness}


A seemingly plausible way to thwart \mia{s} is to add noise to (``perturb") data records before training the model.
The rationale is that \adv is likely to fail at identifying a membership privacy of data record because \adv cannot know what perturbation was added to that record. 

We divided the original training set (``No Noise'') into two subsets of equal size: 1) a \textit{clean} subset without any noise and 2) a \textit{noisy} subset with perturbed samples.
We crafted FGSM noise~\cite{goodfellow2015explaining}, and tested different values of adversarial noise perturbation budget $\epsilon$ ranging from $1/255$ to $352/255$ (under $\ell_\infty$).

Our hypothesis is that adding noise to training data records would lower the \iment accuracy.
Further, the corresponding \method scores would be lower as the noisy samples are more difficult to learn and contribute negatively to the model utility.
The more noise we add, the lower the \method scores, and the lower the \iment accuracy.

\begin{figure*}[htb]
    \centering
    \subfigure[LOCATION]{\includegraphics[width=0.23\textwidth]{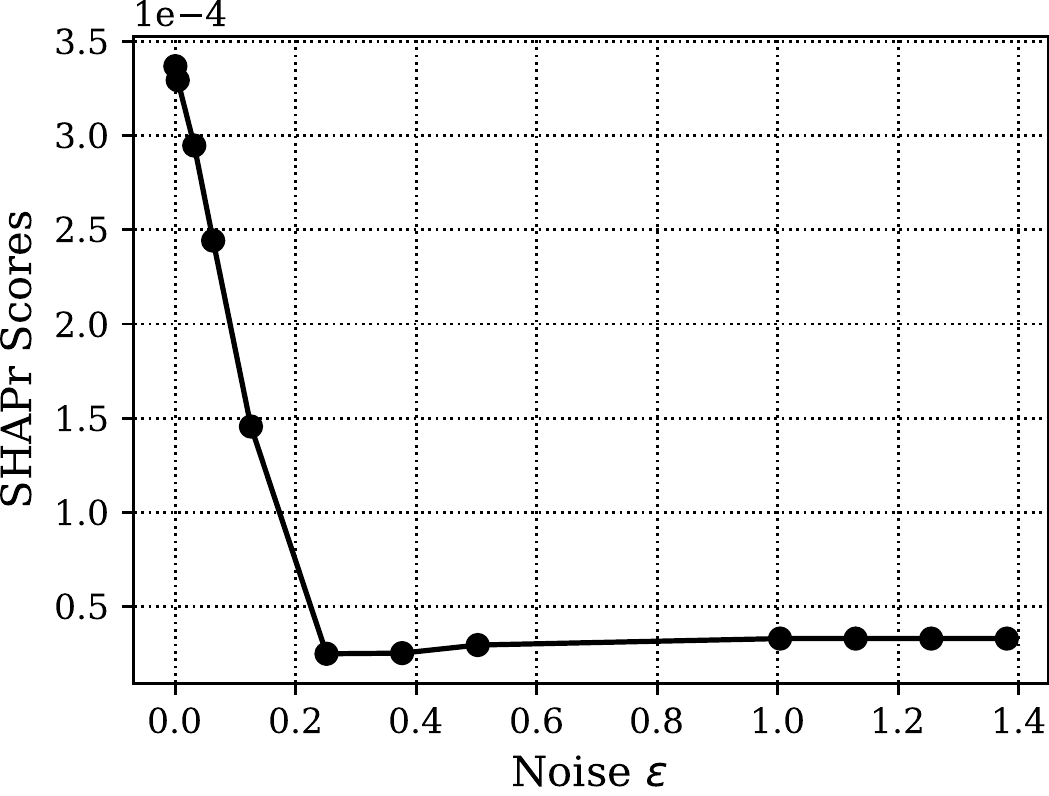}}
    \subfigure[FLOWER]{\includegraphics[width=0.23\textwidth]{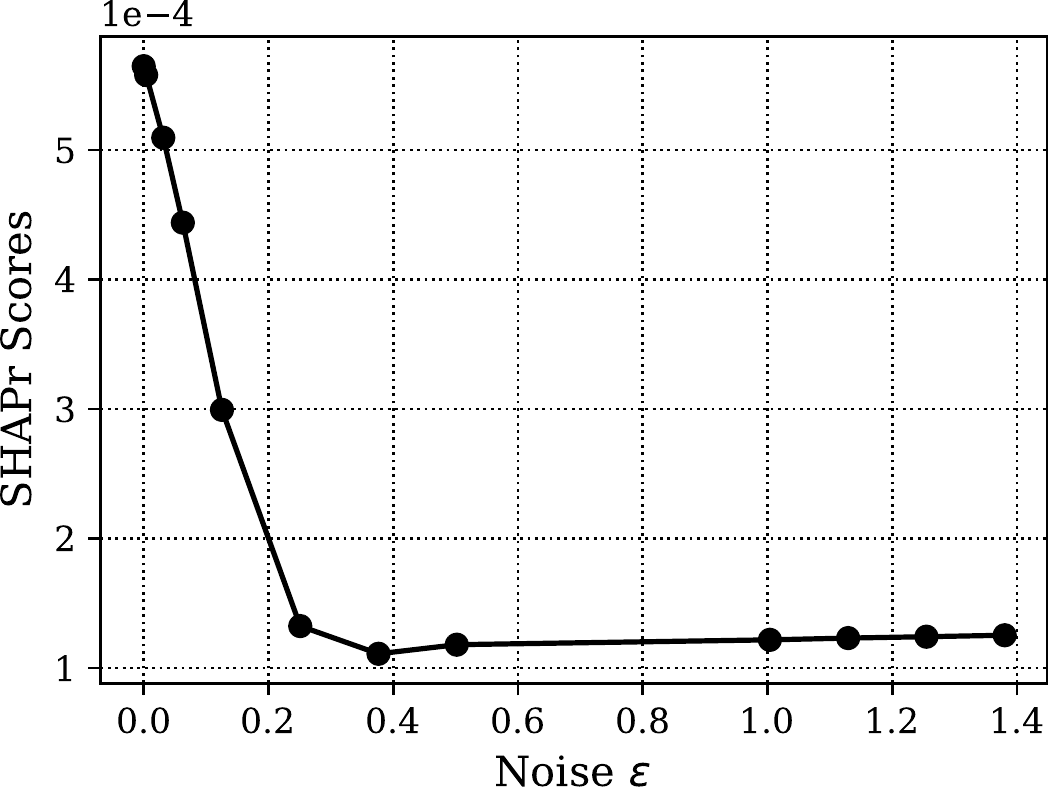}}
    \subfigure[USPS]{\includegraphics[width=0.23\textwidth]{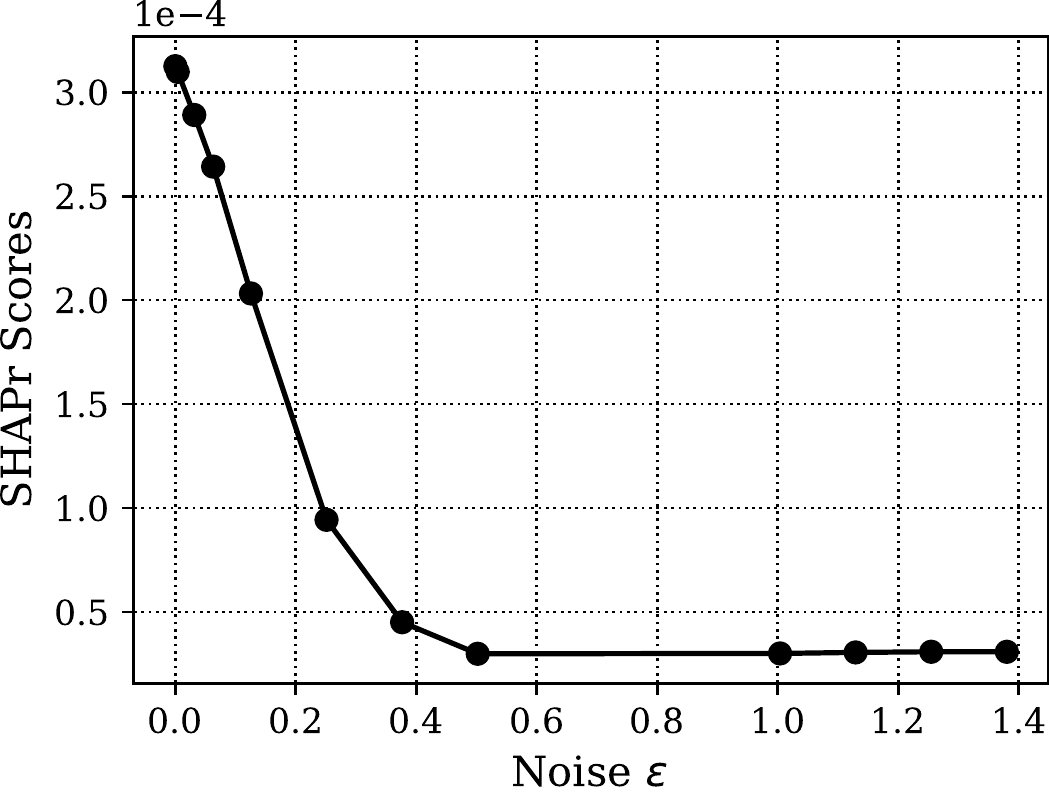}}
    \subfigure[MEPS]{\includegraphics[width=0.23\textwidth]{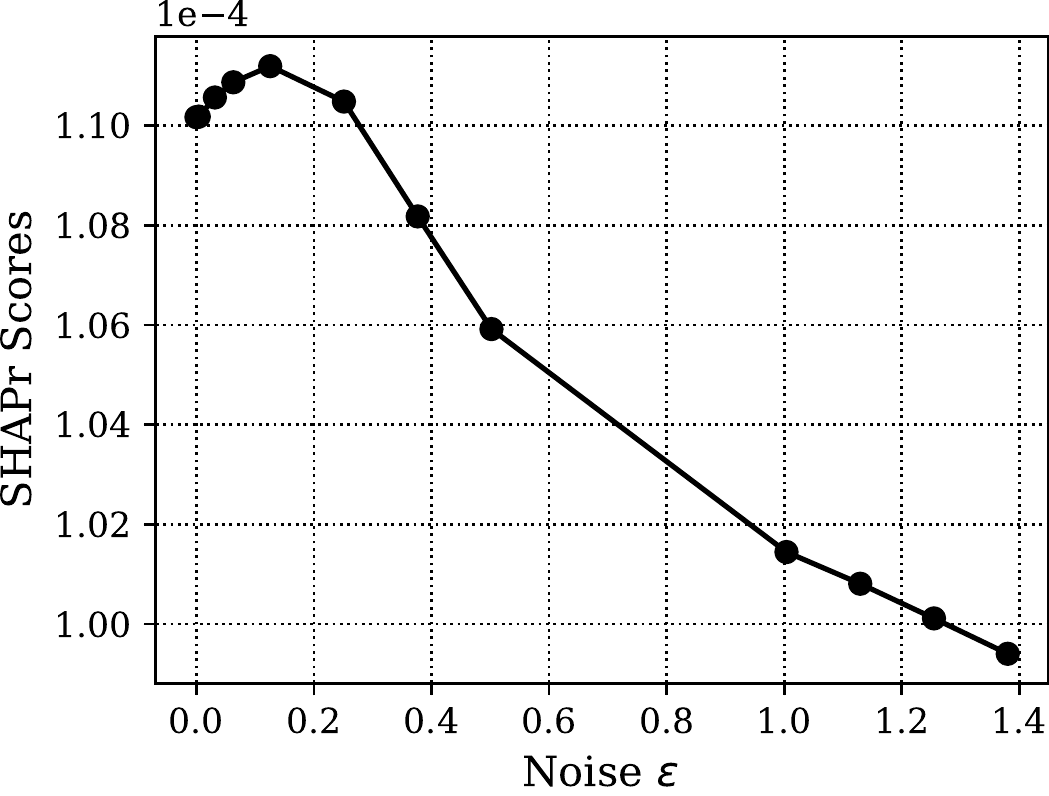}}
    \subfigure[TEXAS]{\includegraphics[width=0.23\textwidth]{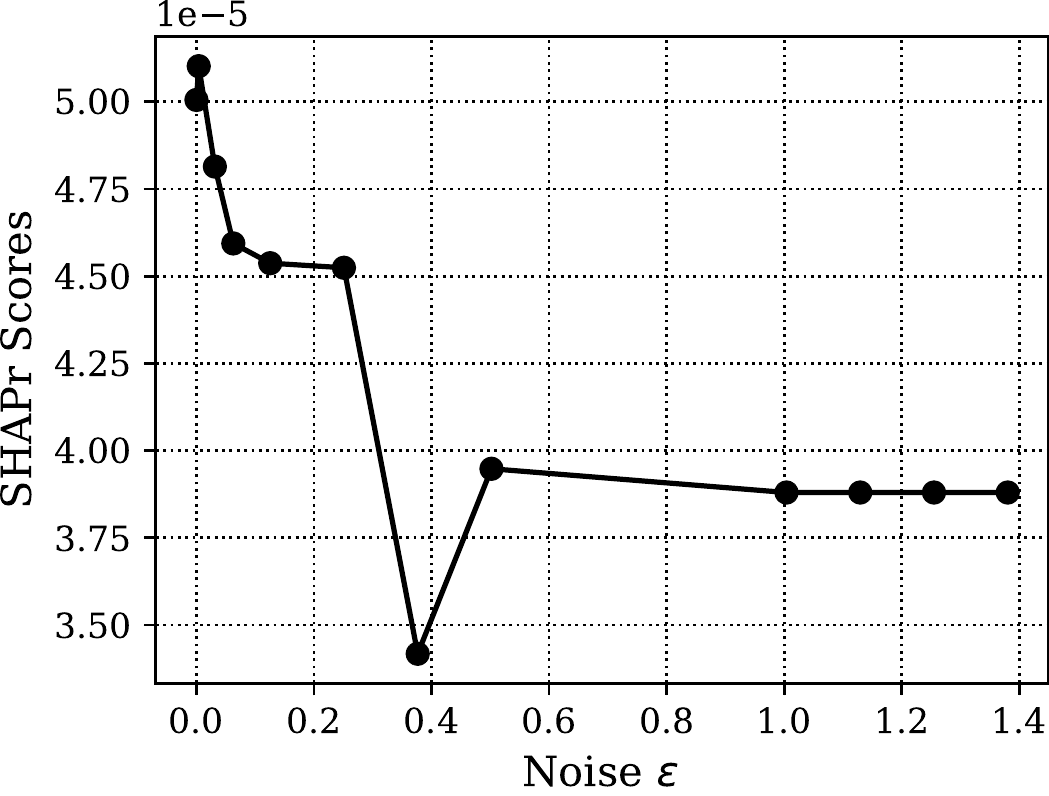}}
    \subfigure[PURCHASE]{\includegraphics[width=0.23\textwidth]{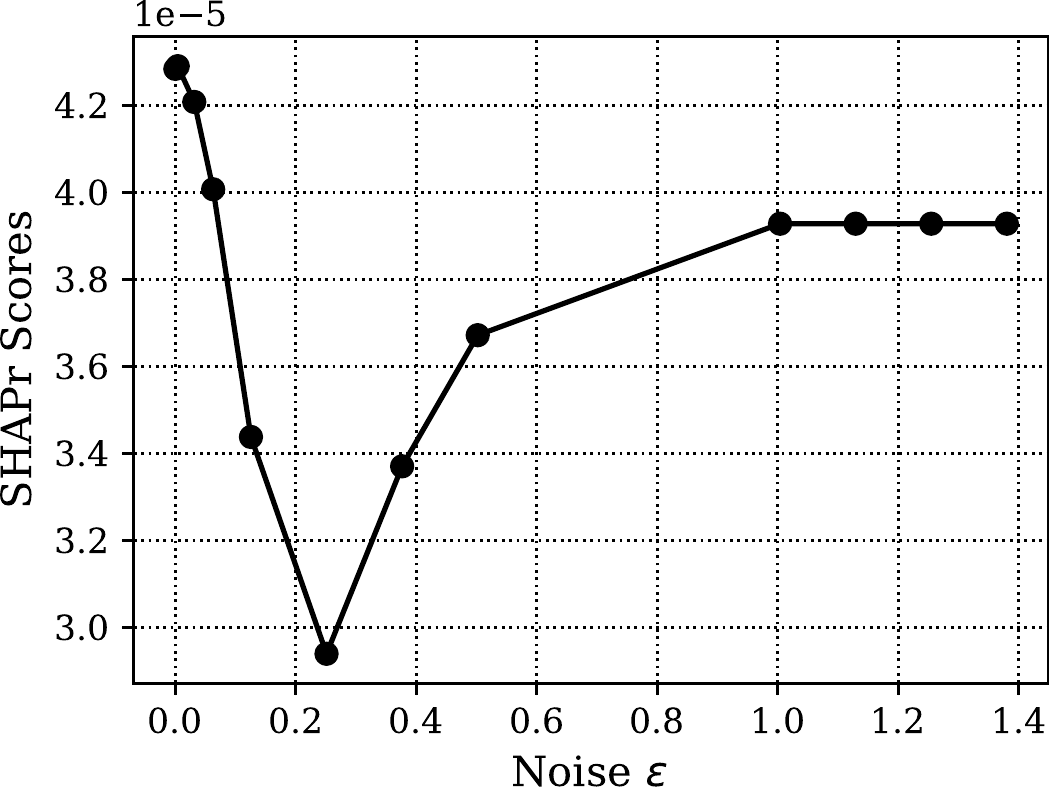}}
    \subfigure[MNIST]{\includegraphics[width=0.23\textwidth]{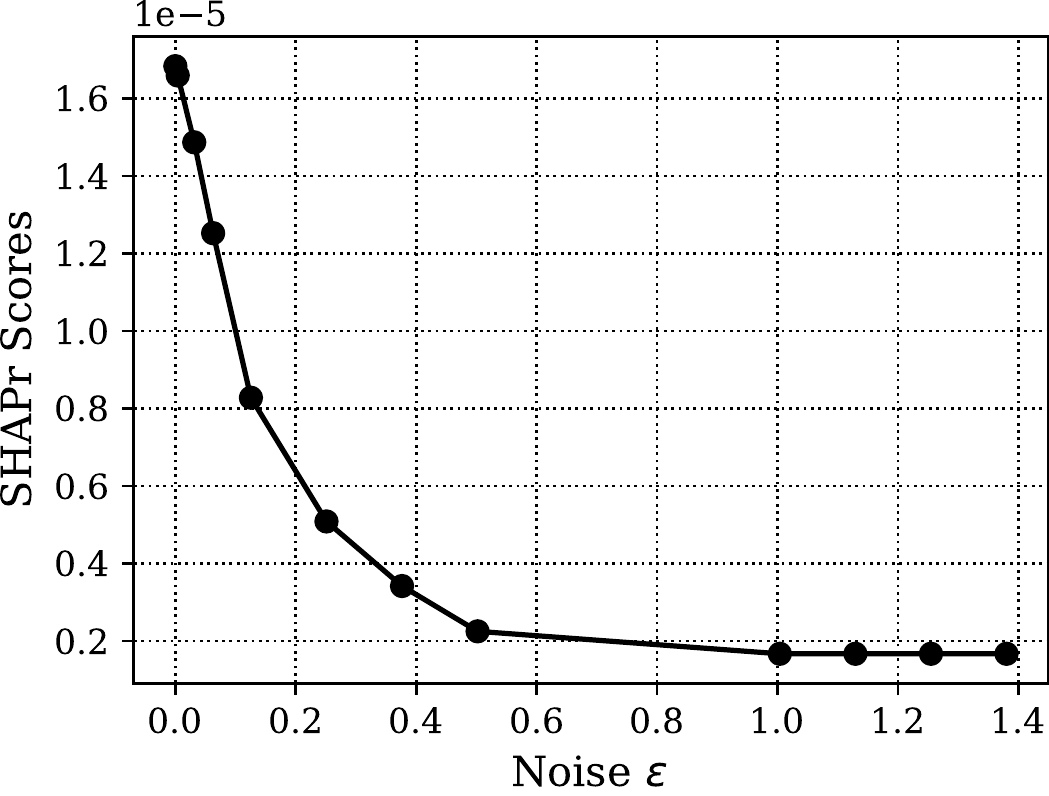}}
    \subfigure[FMNIST]{\includegraphics[width=0.23\textwidth]{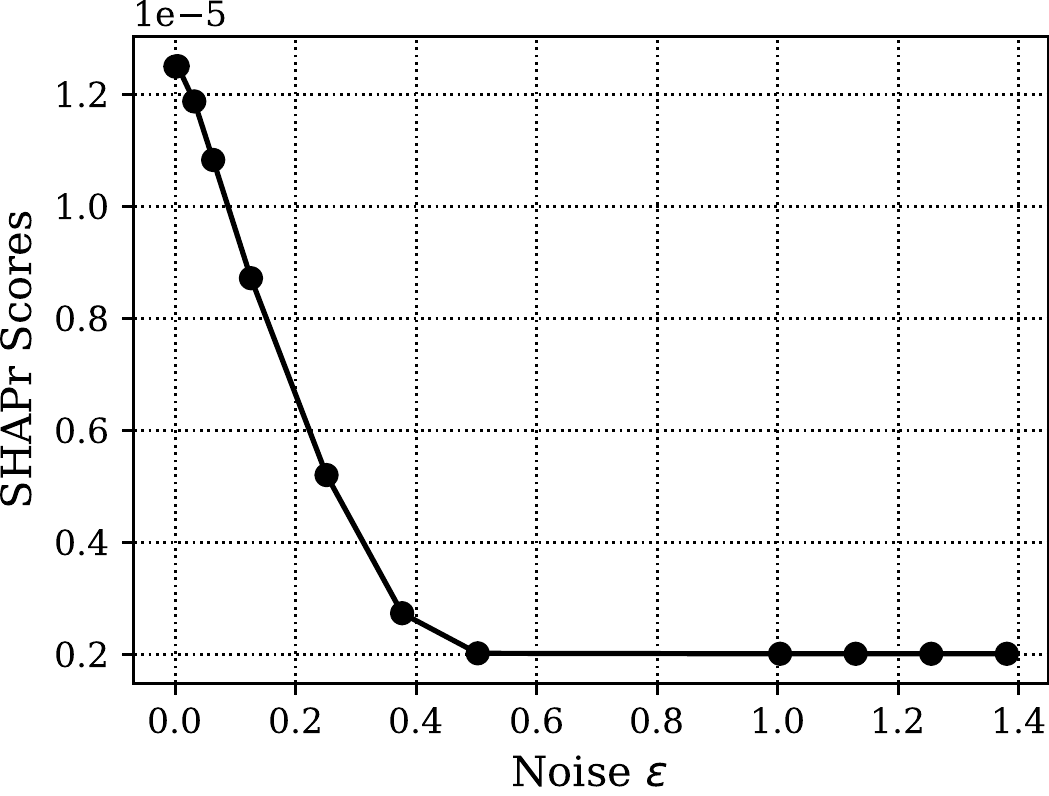}}
    \subfigure[CREDIT]{\includegraphics[width=0.23\textwidth]{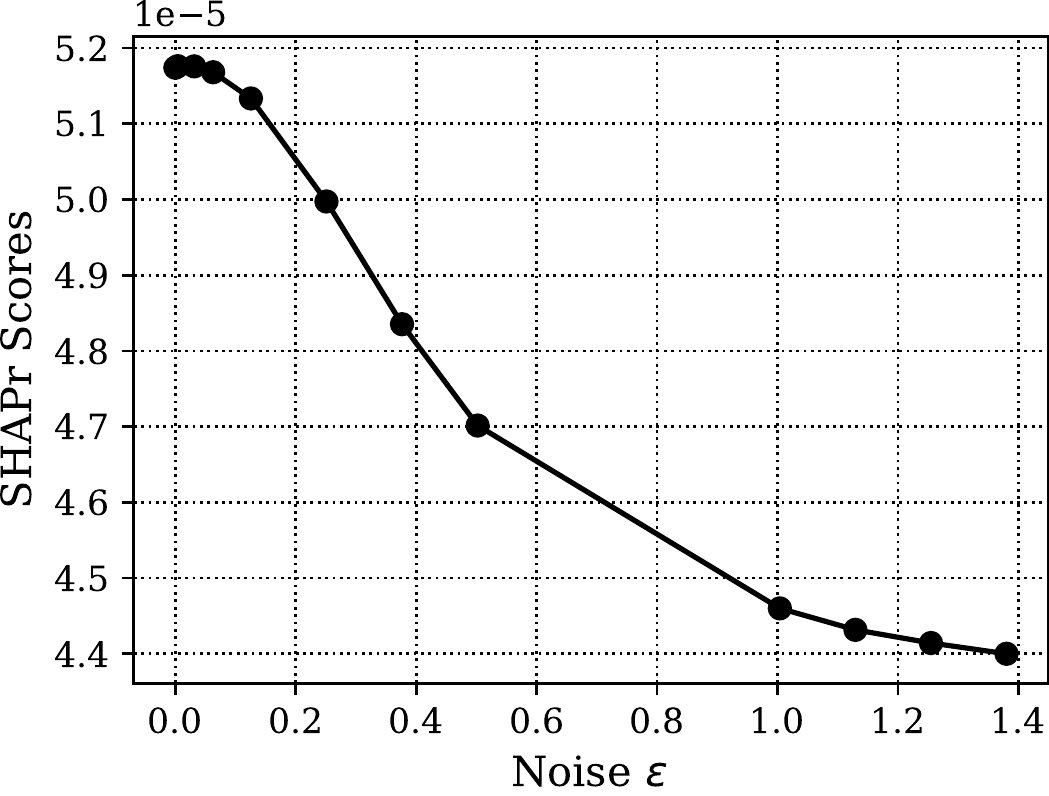}}
    \subfigure[CENSUS]{\includegraphics[width=0.23\textwidth]{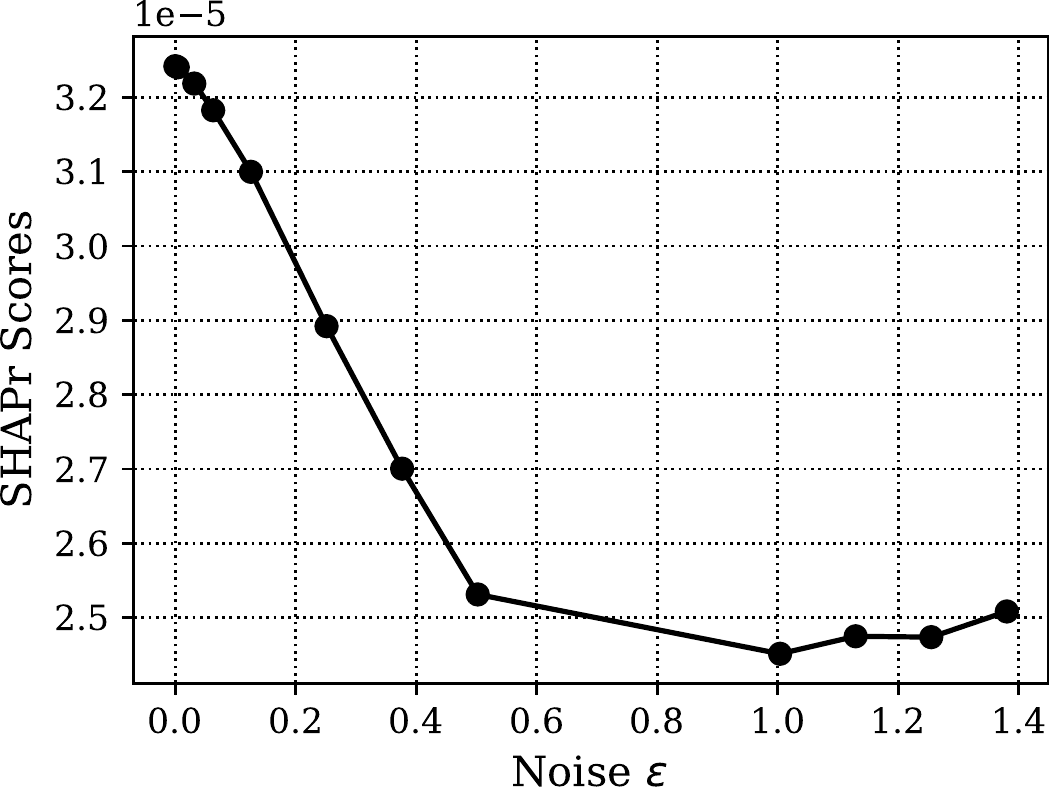}}
    \caption{Adding noise to training data records can lower their susceptibility to \mia{s}.}
    \label{fig:shaprnoise}
\end{figure*}

In Figures~\ref{fig:shaprnoise}, we see that the hypothesis is true:  \method scores decrease on increasing the noise, indicating a decrease in the privacy risk for the noisy data records to \mia{s}. 
\method scores are fine-grained and heterogeneous (Property~\ref{prop4}) which make them sensitive to noise added to the training data records.

Additionally, we use Pearson's correlation coefficient to measure whether \method and \song match the trend in ground truth \mia accuracy. While we note that \method has a positive correlation coefficient across all datasets (Table~\ref{tab:noisecorr_shapley}), \song does not match the trend in ground truth \mia for the noisy data subset. The average score for \song is not impacted by the added noise indicated by several negative correlations (\colorbox{red!25}{red}).
We observe that there is no consistent correlation between \song and \iment accuracy.

\setlength\tabcolsep{2pt}
\begin{table}[h]
\caption{\method correlates with \iment accuracy on noisy subset as seen by the positive Pearson's Correlation Coefficient (referred as ``PCC''), i.e., \method scores follows the decrease in \iment accuracy. \colorbox{orange!25}{orange} indicates that correlation is not significant; \colorbox{red!25}{red} indicates that the correlation is significant and negative 3); and \colorbox{green!25}{green} indicates that correlation is positive and significant.}
\centering
\footnotesize
\begin{tabular}{ |c| >{\centering\arraybackslash}p{2cm}| >{\centering\arraybackslash}p{2cm}| } 
\hline
 \textbf{Dataset} & \textbf{\method PCC} & \textbf{\song PCC}  \\ 
 \hline
 \multicolumn{3}{|c|}{\textbf{\song Datasets}} \\
 \hline
  \textbf{LOCATION} & \cellcolor{green!25}0.89 & \cellcolor{red!25}-0.98 \\ 
 \textbf{PURCHASE} & \cellcolor{orange!25}0.07 & \cellcolor{red!25}-0.58\\
 \textbf{TEXAS} & \cellcolor{green!25}0.84 & \cellcolor{green!25}0.68\\
 \hline
 \multicolumn{3}{|c|}{\textbf{Additional Datasets}}\\
 \hline
\textbf{MNIST} & \cellcolor{green!25}0.60 & \cellcolor{orange!25}0.02\\
\textbf{FMNIST} & \cellcolor{green!25}0.97  & \cellcolor{red!25}-0.65\\
\textbf{USPS} & \cellcolor{green!25}0.43 & \cellcolor{red!25}-0.90\\
\textbf{FLOWER} & \cellcolor{green!25}0.94 & \cellcolor{red!25}-0.90\\
\textbf{MEPS} & \cellcolor{green!25}0.86 & \cellcolor{red!25}-0.88\\
\textbf{CREDIT} & \cellcolor{green!25}0.93 & \cellcolor{red!25}-0.85 \\
\textbf{CENSUS} & \cellcolor{green!25}0.97 & \cellcolor{red!25}-0.80 \\
\hline
\end{tabular}
\label{tab:noisecorr_shapley}
\end{table}

The lack of sensitivity of \song to training data noise can be attributed to clustering of \song around $0.5$ indicating indecisive membership resulting in lack of heterogeneity (Property~\ref{prop4}) as seen Figure~\ref{fig:attplots1} and~\ref{fig:attplots2} for \song's distribution.

\begin{figure}[htb]
    \centering
    \subfigure[Datasets where risk increases]{\includegraphics[width=0.23\textwidth]{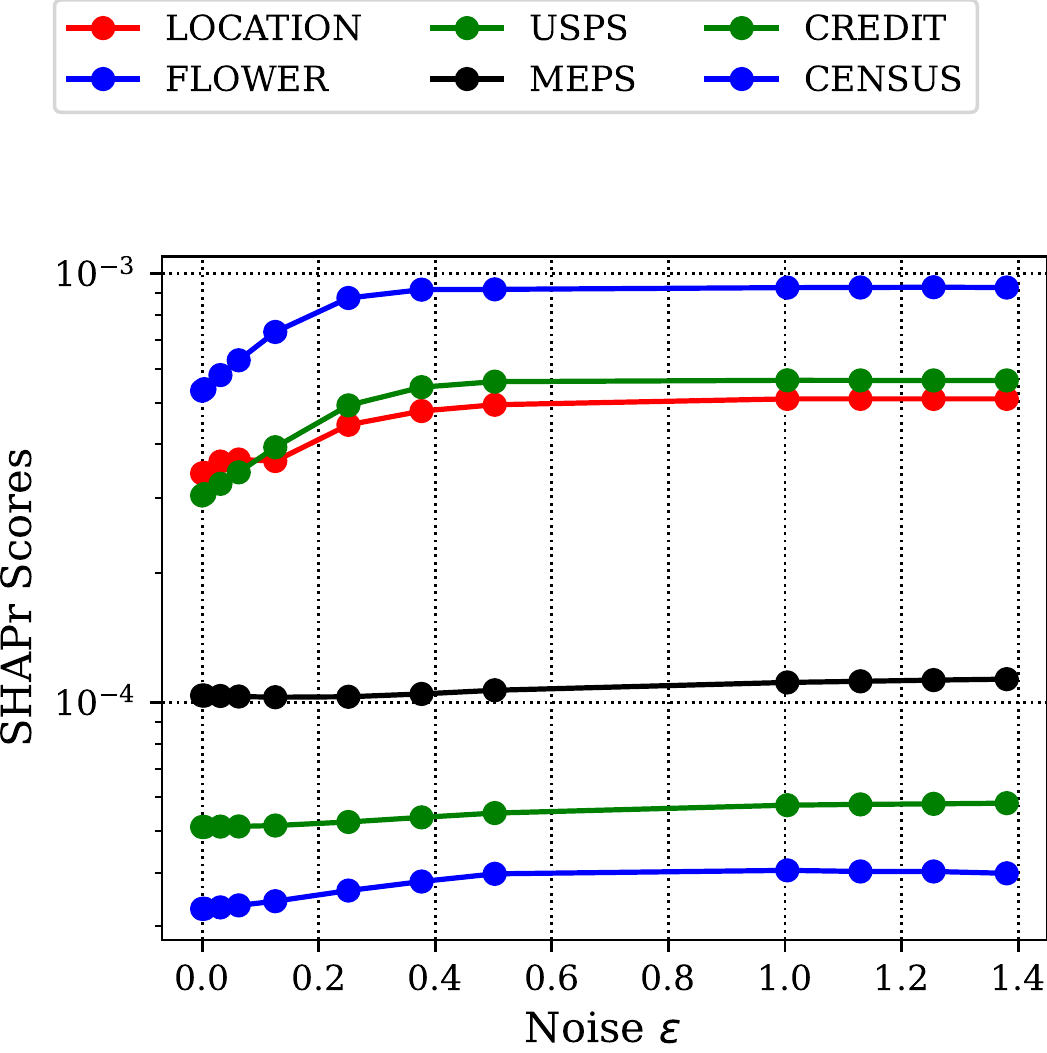}}
    \subfigure[No consistent trend]{\includegraphics[width=0.23\textwidth]{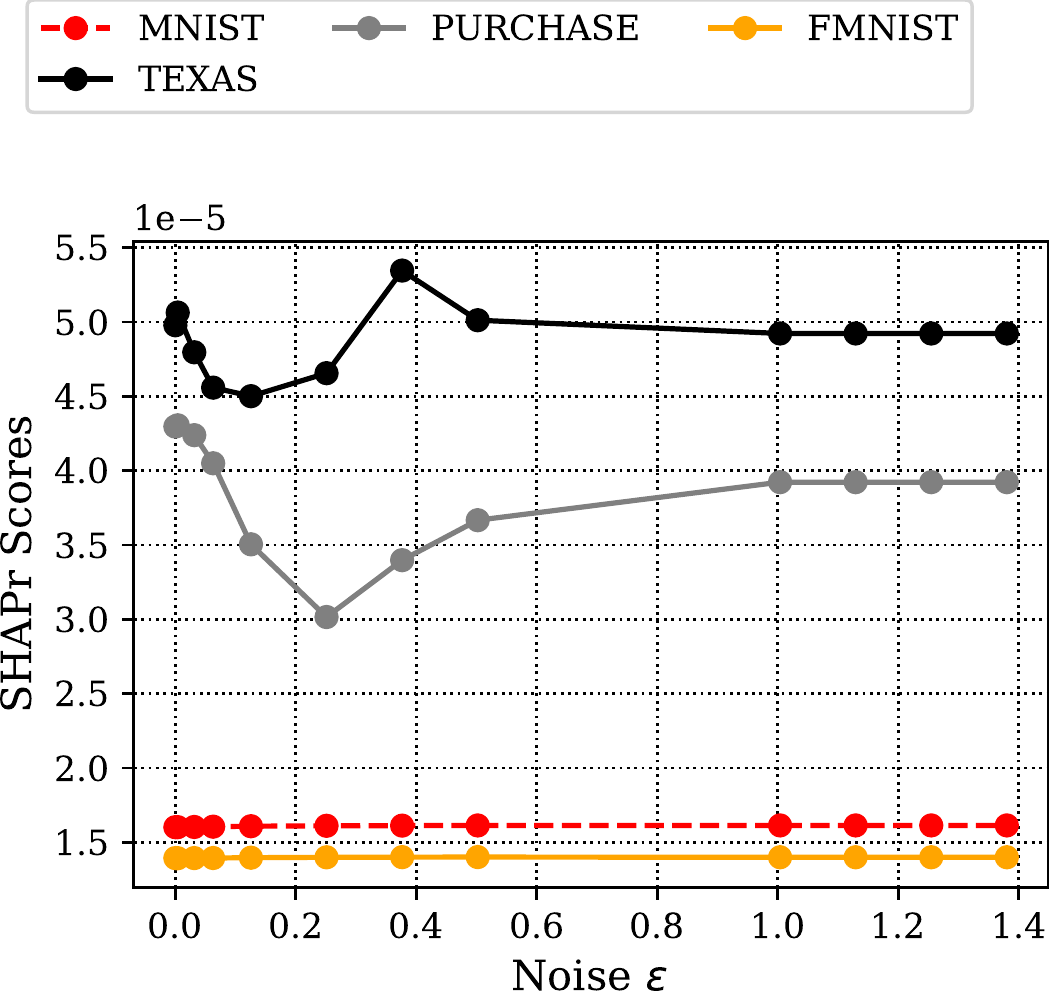}}
    \caption{Adding noise to training data records can lower their susceptibility to \mia{s}.}
    \label{fig:clean_records}
\end{figure}

Having shown that \method can evaluate addition of noise to training data records as a defence, we want to see if this is an effective metric. In some of the datasets: LOCATION, USPS, FLOWER, MEPS, CREDIT and CENSUS, we note that the clean data points in $\mathcal{D}_{aux}$ become more vulnerable to \mia{s} as they become more influential to the utility of the model (Figure~\ref{fig:clean_records} (a)). For some datasets, \method scores do not show a consistent trend (Figure~\ref{fig:clean_records} (b)). We leave the detailed exploration for the reasons behind this for future work.

\end{document}